\documentclass[journal]{IEEEtran}
\usepackage{newcent}
\usepackage{graphicx, color}
\usepackage{amsmath, amssymb}
\usepackage{subfig}
\usepackage{soul}
\usepackage{url}
\usepackage{wasysym}
\expandafter\def\expandafter\UrlBreaks\expandafter{\UrlBreaks
  \do\a\do\b\do\c\do\d\do\e\do\f\do\g\do\h\do\i\do\j%
  \do\k\do\l\do\m\do\n\do\o\do\p\do\q\do\r\do\s\do\t%
  \do\u\do\v\do\w\do\x\do\y\do\z\do\A\do\B\do\C\do\D%
  \do\E\do\F\do\G\do\H\do\I\do\J\do\K\do\L\do\M\do\N%
  \do\O\do\P\do\Q\do\R\do\S\do\T\do\U\do\V\do\W\do\X%
  \do\Y\do\Z}
\usepackage{hyperref}
\usepackage{booktabs}
\usepackage{multirow}
\usepackage{cite}
\usepackage[]{algorithm2e}
\definecolor{mygreen}{rgb}{0,0.6,0}
\usepackage{listings}
\begin{document}
\title{\emph{SDN Partitioning}: A Centralized Control Plane for Distributed Routing Protocols}

\author{
\IEEEauthorblockN{Marcel Caria, Admela Jukan, and Marco Hoffmann}

\thanks{M. Caria and A. Jukan are with the Technische Universit\"at Carolo-Wilhelmina zu
Braunschweig, 38106 Braunschweig, Germany, e-mail:  \{m.caria, a.jukan\}@tu-bs.de}
\thanks{M. Hoffmann is with Nokia Bell Labs, 81541 Munich, Germany, e-mail: marco.hoffmann@nokia.com}
}

\markboth{Preliminary Version / Preprint}%
{Caria \MakeLowercase{\textit{et al.}}: SDN-Partitioned OSPF Domains}

\maketitle

\begin{abstract}
Hybrid IP networks that use both control paradigms -- distributed and centralized -- promise the best of two worlds: programmability and agility of SDN, and reliability and fault tolerance of distributed routing protocols like OSPF. The common approaches follow a division of labor concept, where SDN controls prioritized traffic and OSPF assures care-free operation of best effort traffic. We propose \emph{SDN Partitioning}, which establishes centralized control over the distributed routing protocol by partitioning the topology into sub-domains with SDN-enabled border nodes, such that OSPF's routing updates have to traverse SDN border nodes to reach neighboring sub-domains. This allows the central controller to modify how sub-domains view one another, which in turn allows to steer inter-sub-domain traffic. The degree of dynamic control against simplicity of OSPF can be trade off by adjusting the size of the sub-domains. This paper explains the technical requirements, presents a novel scheme for balanced topology partitioning, and provides the models for common network management tasks. Our performance evaluation shows that -- already in its minimum configuration with two sub-domains -- SDN Partitioning provides significant improvements in all respects compared to legacy routing protocols, whereas smaller sub-domains provide network control capabilities comparable to full SDN deployment.
\end{abstract}


\section{Introduction}
\IEEEPARstart{T}{he} term \emph{hybrid control plane} refers to an increasingly important network architecture, where both control plane paradigms -- the logically centralized Software-Defined Networking (SDN) and a distributed routing protocol like Open Shortest Path First (OSPF) or Intermediate System to Intermediate System (IS-IS) -- are deployed in the same routing domain~\cite{brocade, picos}. While the discussion on centralized \emph{versus} distributed network control planes (e.g., \cite{Pepelnjak, Dixon}) is lively and ongoing in the networking community, the need for a \emph{hybrid} networking paradigm, which can combine the advantages of both, has been broadly recognized~\cite{hybrid_1, hybrid_2, hybrid_3, hybrid_4, Brockners, Vissicchio2}, not least as it provides the only pragmatic migration path to SDN without the expensive replacement of all legacy equipment. Moreover, such an architecture allows for a smooth and total cost of ownership optimized migration to SDN. In fact, many new Internet routers are equipped with an interface for OpenFlow (which is the de facto messaging standard between network devices and SDN controllers) and support a hybrid OpenFlow/OSPF mode. Hybrid control plane architectures typically use the distributed legacy routing protocol for best effort packet forwarding, while the SDN controller injects high priority rules on top for advanced routing configurations.

\par A typical hybrid SDN operation follows a "ships-passing-in-the-night'' strategy, whereby distributed legacy routing and SDN control paradigms are oblivious to what the other one configures. This is known to create a number of challenges in an operational network, including those related to network failures, size of forwarding tables, routing convergence time, thus impeding the chances for SDN to be deployed in carrier networks. In case of network failures, for instance, the uncorrelated control planes may cause forwarding anomalies, like routing loops and black holes~\cite{Vissicchio2}. The size of the router's forwarding information base (FIB) is also an issue, as routers use ternary content-addressable memory (TCAM) to perform memory lookups in one clock cycle, which has to be dimensioned economically due to cost, power consumption, and the required silicon space~\cite{tcam}. A hybrid router, in fact, contains both the OSPF \emph{and} OpenFlow forwarding tables, which increases the required FIB size significantly. Finally, hybrid SDN networks require optimization of the SDN router location, or else their advantages become limited.

\par To address the challenges of a hybrid SDN control plane, we propose \emph{Centrally Partitioned Distributed Routing Domains} as the operational mode and new architecture for hybrid networks, or for short \emph{SDN Partitioning}. In our approach, SDN switches are used as border nodes to \emph{partition} the original distributed (e.g., OSPF) routing domain into sub-domains. With OSPF, for instance, the SDN nodes \emph{appear} to their legacy neighbors as regular OSPF routers, while they actually act as simple protocol repeaters that forward all OSPF messages to the centralized SDN controller, where protocol messages can be modified before they are returned to the sending node and then flooded across the sub-domain border. This in turn allows to reconfigure the routing of traffic \emph{between} sub-domains by determining the exit border node on a per-destination basis. In our scheme, the distributed routing protocol remains stable at all times, while inter-sub-domain routes (which contribute the majority of traffic) are controlled in a centralized fashion. This paper details the network architecture with SDN Partitioning, and provides the complete mathematical background of the scheme, including the theory and complexity of LSA generation and the optimization models for typical network management tasks (i.e., traffic engineering, capacity planning, and fault recovery) as well as the used network partitioning method that generalizes a prominent model for the vertex separator problem. Our numerical analysis shows that, in all evaluated measurements, the performance of SDN Partitioning ranges from significant improvements to regular OSPF (in its minimum configuration with only two sub-domains), up to network control capabilities comparable to full SDN deployment (with a partitioning in into smaller sub-domains).

\par The rest of the paper is organized as follows: Section~\ref{relatedwork} discusses the related work. Section~\ref{arch-section} presents the technological background and details the assumed network architecture. Section~\ref{math_section} provides the network model, the mathematical background for the generation of customized routing updates, and the used graph partitioning approach. The required optimization models for common network management tasks are explained in Section~\ref{optimization}. Our numerical evaluation is presented in Section~\ref{results} and Section~\ref{conclusions} concludes the paper.

\section{Related Work and Our Contribution}\label{relatedwork}

\par Hybrid SDN networking has been analyzed and explained to a great extent in~\cite{hybrid_1, hybrid_2, hybrid_3, hybrid_4, tamal_ICC}. In a hybrid network, the capability of the SDN controller to insert higher priority rules into the forwarding tables is a powerful new feature which in~\cite{steroids} has  been coined as ``policy based routing on steroids''. We therefore refer to this control plane approach in the performance analysis of this paper (Section~\ref{results}) as the \emph{stacked hybrid} model. In~\cite{hybrid_2}, it was analyzed how this hybrid approach can be used as an efficient migration strategy for gradual SDN deployment in legacy networks. It was found that for a given network topology and migration planning horizon, the sequence in which IP routers are replaced with SDN-enabled routers has large impact on network performance. A known practical implementation of a hybrid control plane is Google's \emph{B4}~\cite{B4}. Our architecture is different from any previously proposed hybrid SDN control plane architecture, since traffic through legacy routers can be steered dynamically by customized protocol messages from the centralized SDN controller.

\par In regard to the partitioning of the network, our method sets the goals similar to the partitioning of an OSPF domain into areas (defined in RFC 2328~\cite{ospf}), where OSPF areas are used to simplify the administration of large topologies and to reduce the amount of protocol traffic. In addition, SDN Partitioning allows SDN-based traffic engineering by controlling the routes of inter-sub-domain traffic. Also, partitioning a network into \emph{individual OSPF domains} and connecting them with the Border Gateway Protocol (BGP) would lead to degree of freedom regarding routing control almost similar to what our method can offer. In such a network, BGP could be used for load balancing similar to~\cite{bgp}. However, our method does not require the partitioning into multiple \emph{autonomous systems} and provides a clean separation from the BGP setup that a network has already in place. Another approach partitions the network into \emph{zones}, whereby each node belongs to a single zone only~\cite{vissicchio}. Our partitioning idea however is different from the zone approach, as a zone is defined as a set of interconnected nodes controlled by the same paradigm (i.e., SDN \emph{or} OSPF). In contrary, a sub-domain in our approach is a subgraph of OSPF routers, while the SDN nodes also participate in OSPF.

\begin{figure*}[t] \center
\includegraphics[width=\textwidth]{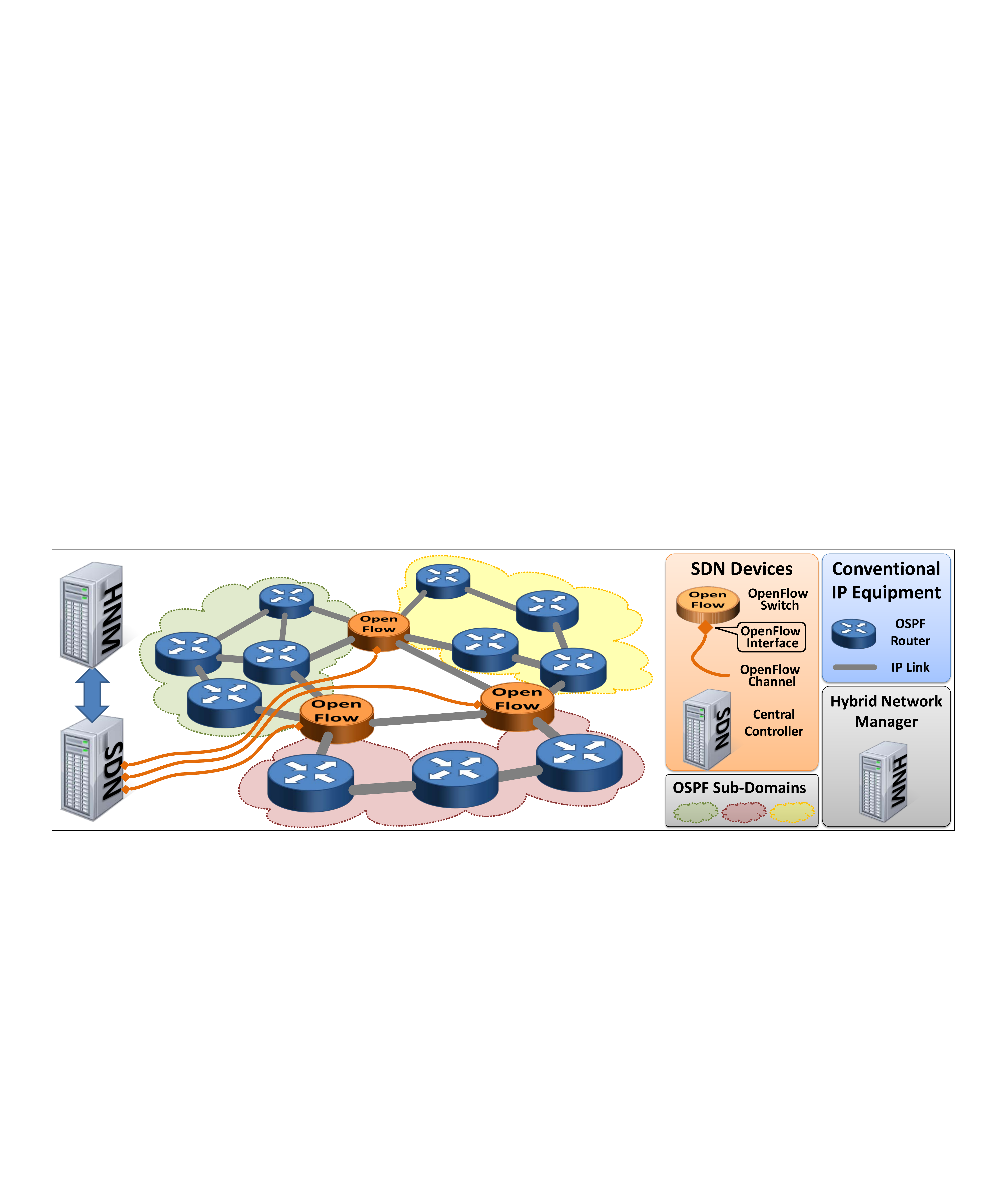}
\caption{The network architecture of an SDN-partitioned OSPF domain.}
\label{arch} \end{figure*}

\par In regard to the routing protocol and architecture, this work has a strong relation to, but a distinctive differences from the line of work published under the name ``Fibbing'' by Vissichio et al. in~\cite{fibbing1} and~\cite{fibbing2}. The similarity of our proposal and this work is in the idea to alternate routing in the network by introducing \emph{fake} information into the legacy routing protocol (e.g., OSPF). However, the mode of operation, the requirements, and the expressiveness in our and the Fibbing schemes are different. While Fibbing requires only to extend the existing network architecture with the so-called Fibbing controller, SDN Partitioning necessitates the deployment of SDN-enabled routers. More in detail, the \emph{operational} difference is as follows: Fibbing floods fake \emph{external} (i.e., Type 5) link-state advertisements (LSAs) through the network to extend the actual topology with virtual nodes and links using the Forwarding Address field in this type of LSA, which in turn let the actual OSPF nodes recompute their shortest paths. SDN Partitioning, on the other hand, uses SDN-enabled routers to interrupt OSPF's flooding mechanism for \emph{all} LSAs at sub-domain borders to allow the continuation of the same flooding process with customized (i.e., optimized) LSAs in the neighboring sub-domain. Though both schemes are limited by OSPF's destination-based forwarding behavior, Fibbing is slightly more expressive than SDN Partitioning, as it provides \emph{full} control over any destination's next hop at any router, whereas SDN Partitioning preserves OSPF's control over locally limited (i.e., intra-sub-domain) traffic. By design, Fibbing uses the Forwarding Address field in Type 5 LSAs, which is considered as an ``exotic feature'' of OSPF (e.g., see Cisco's~\cite{ciscoproblem} and Juniper's\cite{juniperproblem} knowledge base articles, and~\cite{OSPFforwardingaddress}). Also, Fibbing depends on external traffic measurement tools to allow for routing optimizations, whereas SDN Partitioning is self-contained in this regard: all flows across SDN-enabled routers are automatically monitored by the central controller. The later feature is standard to SDN, since OpenFlow uses byte counters for all flow table entries. Thus, a routing optimizer does not need external tools and can use Simple Network Management Protocol (SNMP) link counters from the inside of the sub-domains in addition to the flow counters. Finally, SDN Partitioning \emph{decreases} the protocol overhead in the network by simply suppressing those LSAs at border nodes that are without an effect for the particular sub-domain, which is not the case in Fibbing.

\par We note that in this paper, we interchangeably use the terms SDN and OpenFlow, whereas the latter actually denotes a protocol (and the de facto standard) for the communication between the centralized controller and SDN-enabled devices. Secondly, we also interchangeably use the terms OSPF and distributed routing protocol in this paper, whereas in this case, the former actually denotes a particular variety of the latter. IS-IS, which is like OSPF a so called \emph{link state routing protocol}, applies for all intents and purposes in this paper. The Routing Information Protocol (RIP), on the other hand, is a \emph{distance vector routing protocol}, which is not flooding topology information, and cannot be used for SDN Partitioning.

\subsection*{Our contribution}
This paper is an major extension of our previous work~\cite{divideandconquer} and~\cite{caria_HPSR}. In~\cite{divideandconquer}, we studied traffic engineering in SDN Partitioned networks and showed that relatively few SDN-capable nodes are needed as compared to full SDN deployment for the same traffic engineering objectives, while~\cite{caria_HPSR} used SDN Partitioning to incorporate dynamic optical circuit provisioning (also referred to as \emph{optical bypass}) into the operation of the load balancer. The mathematical models proposed in these papers were the basis for the new load balancing model that we present here in Subsection~\ref{math_te}. The model developed here has less variables and uses more pre-computed parameters, which allows the optimization of larger topologies. Furthermore, we provide in this paper two additional optimization models, one for capacity dimensioning and one for efficient network fault restoration. We also include more comprehensive systems and implementation considerations, including a theoretical analysis of LSA generation. Finally, while our previous work used a brute force approach to partition networks, this paper applies a new and efficient method to find balanced vertex separators. This method generalizes the integer linear programming (ILP) formulation of the vertex separator problem published by Balas and de Souza in~\cite{Souza}, which in our version allows for the partitioning of a graph into an arbitrary number of subgraphs with a surprisingly good performance. The graph-theoretical aspects of network partitioning, however, are for the most part out of scope for the present paper, as we here use it as a tool to analyze the presented hybrid control plane scheme.

\section{Background}\label{arch-section}
Figure~\ref{arch} illustrates the idea of SDN-based partitioning: SDN-enabled Internet routers replace legacy routers at some strategic locations. As it is well known, mesh topologies can be partitioned in various ways: the more SDN routers in the network the larger the number of sub-domains. We will go into details of network partitioning in the next section. Let us assume for the time being that the best possible partitioning method was used and as a result the distributed routing domain in Figure~\ref{arch} has been partitioned in three sub-domains: out of 14 legacy routers, three were replaced by SDN enabled routers, and the rest of the legacy routers are now associated with the newly created distributed routing sub-domains. 

\par The network deploys conventional IP routers that run OSPF, while three nodes at specific locations have been replaced with OpenFlow switches that establish individual control channels to the SDN controller through their OpenFlow interfaces, and then act as border nodes to the OSPF partitions (or, sub-domains). Note that in our network we do not deploy typical \emph{hybrid routers}, i.e., capable of both OSPF and OpenFlow simultaneously (like \cite{brocade}). Instead, these are standard OpenFlow switches. The compatibility of SDN switches with legacy routers in our operational scheme is provided by the fact that SDN switches \emph{act} as legacy routers and respond with OSPF-conform protocol messages, as generated by the SDN controller. All received OSPF routing protocol messages are -- instead of being flooded to the opposite sub-domain -- transferred to the central SDN controller, which in turn uses this information to enrich its own global view on the current state of the network. The response messages are in fact not generated by the SDN switches (which act as packet forwarding switches), but by the SDN controller, before flooding is continued on the other side.

\par In this architecture, legacy routers are neither aware of the existence of the hybrid control plane, nor of the fact that they belong to some sub-domain. Moreover, all legacy routers connecting directly to SDN border nodes do not recognize any difference to normal OSPF routers. Across these particular links, the \emph{OSPF neighbor adjacency}\footnote{Directly connected OSPF routers use OSPF's Hello protocol to form neighbor adjacencies, which is the prerequisite for any further protocol interaction.} is actually formed between the OSPF router and the central SDN controller \emph{through} the SDN switch. This mechanism remains transparent for OSPF, such that the SDN switch is nonetheless considered as regular OSPF router by its legacy neighbors. The control over the routing (of inter-sub-domain traffic) can now be achieved through the controller's manipulation of the global topological view. Please note that this operational scheme requires the implementation of the used legacy protocol in the central SDN controller (which to date is not standard), and to form neighbor adjacencies to all legacy nodes adjacent to all SDN nodes. Figure~\ref{arch} also depicts a \emph{Hybrid Network Manager} connecting to the SDN controller, which represents the implementation of various routing optimization schemes (like those detailed in Section~\ref{optimization}) through various functional components, as well as the implementation of regular network management functions (e.g., monitoring, service provisioning, etc.).

\begin{figure}[h] \center
\includegraphics[width=5cm]{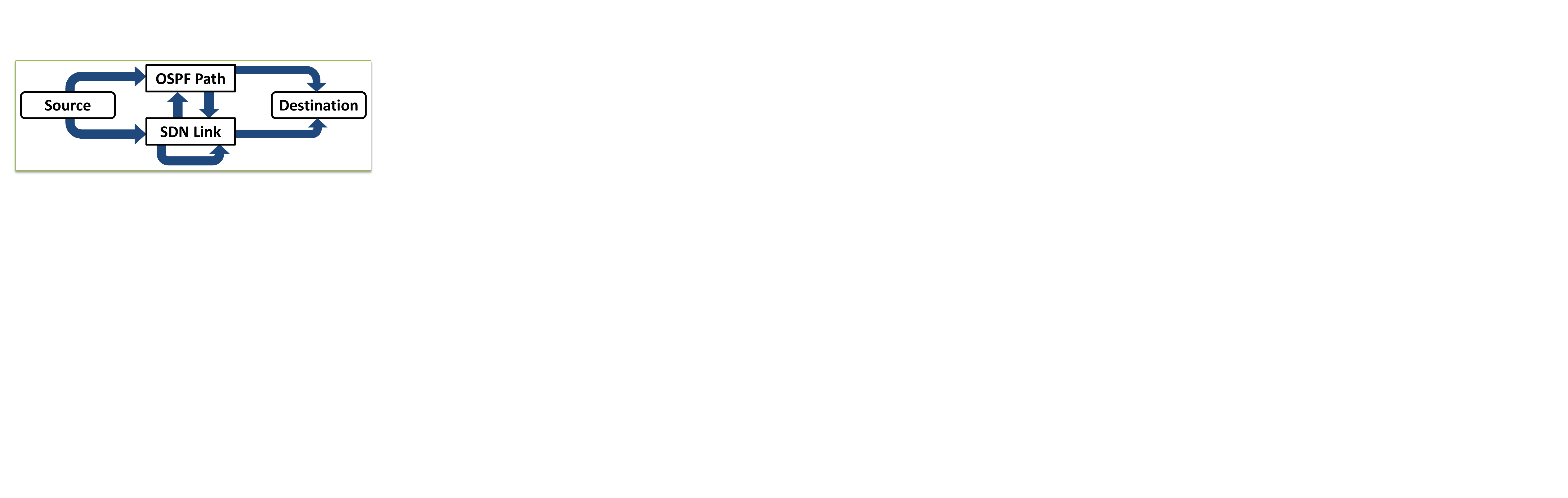}
\caption{Concatenation of path elements.}
\label{flowchart} \end{figure}

\subsection{Routing}\label{routing}

\par In this network architecture, a valid routing path is a concatenation of OSPF paths and SDN links, as depicted Figure~\ref{flowchart}. Note that our particular notion of the terms \emph{OSPF path} and \emph{SDN link} differs from common usage: We define an SDN link as a directional connection between an SDN router and any other (SDN or OSPF) router. An OSPF path is defined as the \emph{unique least cost path} (i.e., protocol mechanisms like Equal Cost Multi Path are not considered) between a (non-SDN) OSPF router and any other (SDN or OSPF) router \emph{within the same sub-domain}.

\par
\begin{figure}[h] \center
\includegraphics[width=8cm]{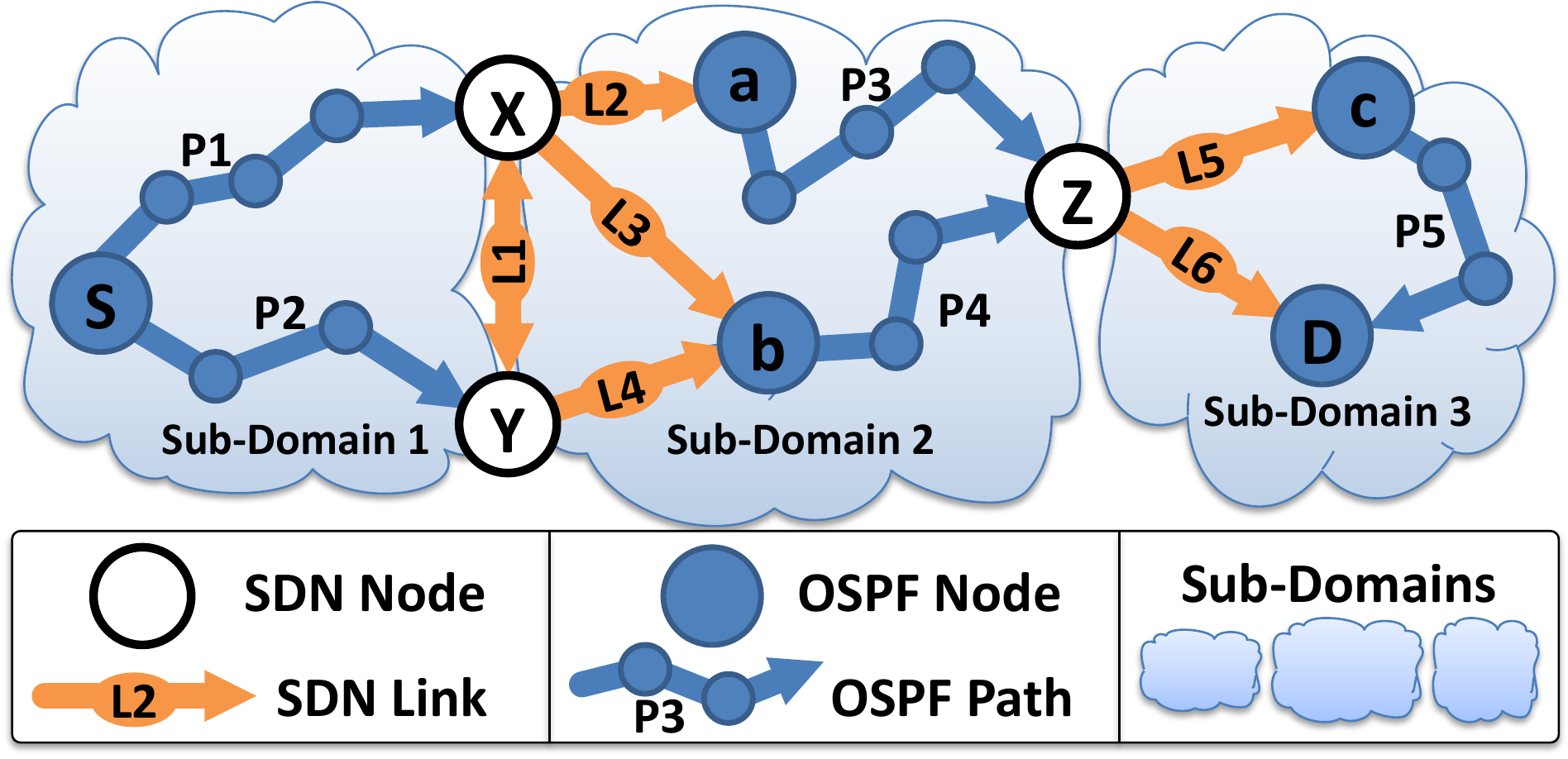}
\caption{Constraints on routing in SDN Partitioning.}
\label{routing_scheme} \end{figure}

 Figure~\ref{routing_scheme} further illustrates the routing constraints. The source node $S$ has exactly one least cost path to each SDN border node ($X$ and $Y$) in Sub-Domain~1. Thus, the route from $S$ to $D$ starts either with OSPF path $P1$ or $P2$, depending on the aggregated cost metric for the routing to $D$, i.e., the aggregated metrics along $P1$ plus the metric advertised by $X$ for reaching $D$, and the aggregated metrics along $P2$ plus the metric advertised by $Y$ for reaching $D$. The next element in the route to $D$ is an SDN link. As an SDN node's flow table can be arbitrarily configured, e.g., for packets matching source addresses $S$ and destination addresses $D$, we see that $P1$ can be continued with SDN links $L1$, $L2$ or $L3$. In fact, $P1$ can be continued even with SDN links $L1$ and $L4$ successively. Note the self loop of the \emph{SDN link} box in Figure~\ref{flowchart}: Because forwarding in SDN nodes is arbitrarily configurable and not constraint by OSPF, SDN links can be concatenated arbitrarily. Arriving at any of the two first OSPF routers ($a$ or $b$) in Sub-Domain~2, there is no choice to be made, because each of the two routers has only a single OSPF path to border node $Z$. Finally, $Z$ can then be configured to forward the packets directly via link $L6$ to $D$, or in case of congestion on that link, forward the packets via $L5$ to OSPF node $c$, from where the packets again have to take the regular OSPF path to the destination.

\subsection{How SDN Partitioning avoids the common mistakes}\label{how_we_avoid_flaws_of_stacked_model}
\par As mentioned earlier, the common mode of operation of a hybrid control plane (which we also referred to as the \emph{stacked} model) requires that a subset of the network nodes are hybrid routers, i.e., routers capable of both OSPF and OpenFlow simultaneously. This operational mode exhibits a number of design flaws. 

\par First, the FIB in a hybrid router is required to hold forwarding entries of OpenFlow \emph{and} OSPF. As the FIB is commonly implemented in hardware to meet the requirements of interface line rates, it uses TCAM that can perform memory lookups in one clock cycle. This type of memory is known to be expensive, power hungry, and demanding in terms of silicon space~\cite{tcam}. As a result, hybrid routers are either provided with weakly dimensioned FIBs or are significantly more expensive than OSPF \emph{or} OpenFlow routers. SDN Partitioning, on the contrary, does not require hybrid routers, and plain OpenFlow switches are sufficient. The OSPF protocol is taken care of in the central controller and the border nodes forward all OSPF messages in a simple repeater mode: incoming messages from OSPF neighbors are forwarded through the OpenFlow channel to the central controller, and vice versa.

\par Second, in case of network failures the two routing systems in the stacked model remain unaware of each other's configurations. As a consequence, failure recovery requires individual and sequential recovery processes, as the SDN part of the network has to wait for the OSPF part to completely converge before it can start its own routing optimization and reconfiguring. To make matters worse, both control planes may jointly cause forwarding anomalies, like routing loops and black holes~\cite{Vissicchio2}, due to individual and mutually inconsistent recovery actions. All of this is not the case in SDN-partitioned networks, as its hybrid control and management system is aware of the network status (topology, routing, and link utilization) in all OSPF sub-domains, as it receives according LSAs in case of a failure instantly. The complete visibility of the network allows our system to pre-calculate actions for network failures to an extent, albeit the challenge would remain in case of multiple failures. 

\begin{figure}[t] \center
\includegraphics[width=8.3cm]{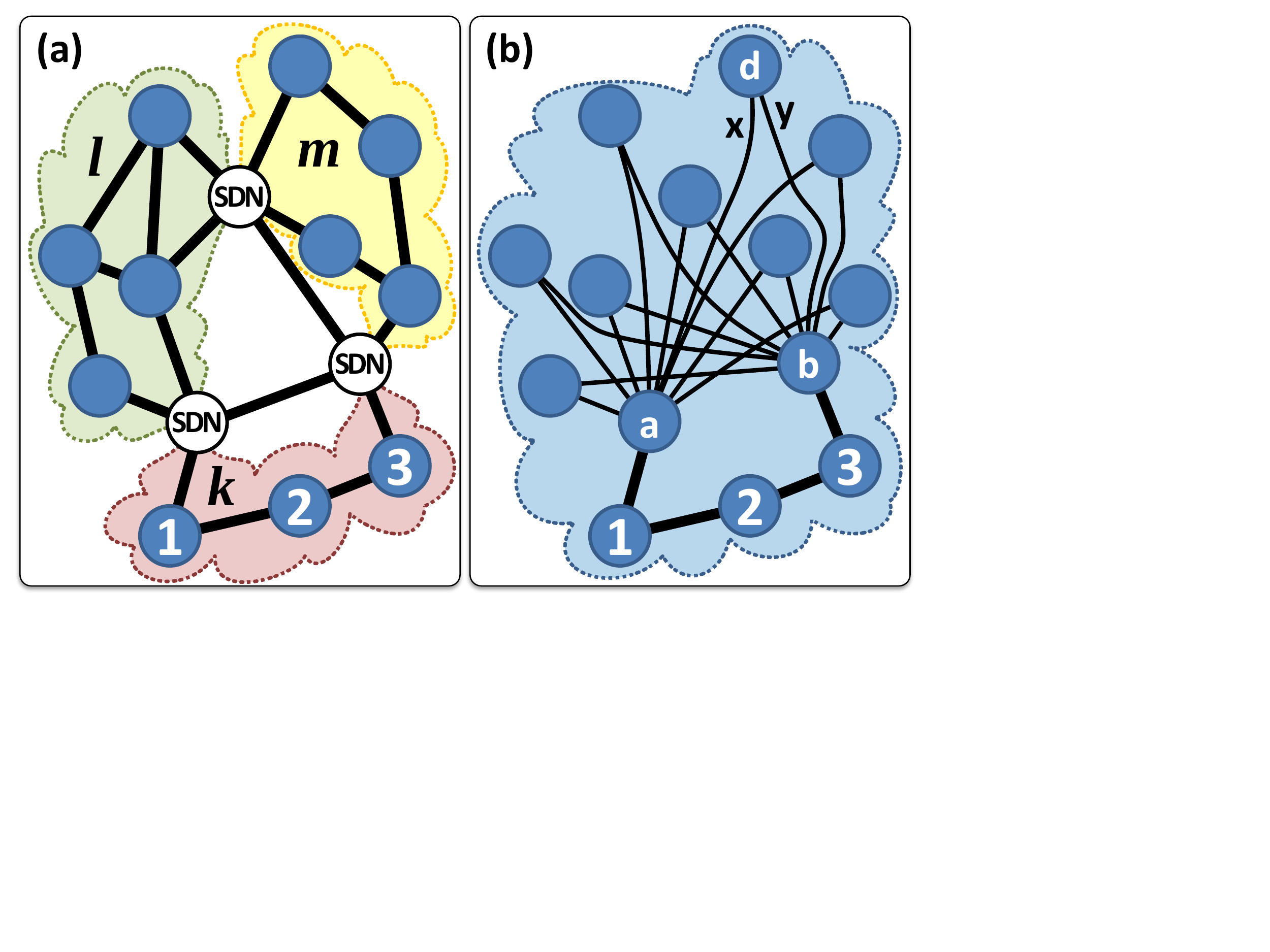}
\caption{Two views of the same network: (a) the actual topology and the partitioning of the network, and (b) how it's represented to the nodes in sub-domain $k$.}
\label{view} \end{figure}

\par Finally, SDN-enabled routers need to be optimally located in a network with a stacked hybrid control plane. Like many other studies, we also have analyzed the SDN switch location problem in~\cite{hybrid_2}, where we also detail that a poor deployment strategy can lead to a significantly lower performance. However, location optimization is questionable under dynamic traffic conditions, as new high priority traffic flows may not pass enough hybrid routers to provide sufficient traffic engineering capabilities. For fairness, SDN-partitioned networks can suffer from the same issue, especially in large sub-domains. However, SDN Partitioning can avoid this pitfall more easily: A flow with a long routing path is most likely traversing multiple sub-domains, and traffic engineering capabilities are provided in a per-sub-domain fashion.

\section{Network model}\label{math_section}
\par This section focuses on the theoretical background, problem complexity, and algorithms of the two fundamental building blocks in our network model: routing modification via LSA generation and network partitioning. The notation used is summarized in Table~\ref{netwsymbls}. 

\begin{table}[h]\begin{center}\footnotesize 
\begin{tabular}{ c l }
\toprule
\textbf{Set} & \multicolumn{1}{c}{\textbf{Meaning}} \\
\midrule
\multirow{2}{*}{$X_k\subseteq X$} & subset $X_k$ denotes all elements \\
 & $\text{ }$ $x\in X$ that belong to sub-domain $k$ \\\addlinespace[1.0mm]
$R$ & set of all sub-domain-internal (OSPF) nodes $r$ \\\addlinespace[1.0mm]
$B$ & set of all border (SDN) nodes $b$ \\\addlinespace[1.0mm]
$N$ & set of all nodes $n$ with $N=R\cup B$ \\\addlinespace[1.0mm]
$\overline{N}_k$ & set of all nodes in $N\backslash (R_k\cup B_k)$ \\\addlinespace[1.0mm]
$[1,K]$ & set of all sub-domains $k$ \\\addlinespace[1.0mm]
$L$ & set of all links $\ell =(n_1,n_2)$ \\\addlinespace[1.0mm]
$F$ & set of all traffic flows $f$ \\\addlinespace[1.0mm]
$P$ & set of all paths $p$ \\\addlinespace[1.0mm]
$M$ & set of all metric vectors $\vec{m}$ \\\addlinespace[1.0mm]
$E$ & set of all exit vectors $\vec{e}$ \\\addlinespace[1.0mm]
$Q$ & set of all quantity vectors $\vec{q}$ \\\addlinespace[1.0mm]
$T$ & set of all link capacity types $t$ \\\addlinespace[1.0mm]
$Y$ & set of all linear cost functions $y$ \\\addlinespace[1.0mm]
\midrule
\textbf{Integer} & \multicolumn{1}{c}{\textbf{Meaning}} \\
\midrule
$m(n_1,n_2)$ & metric of link $(n_1,n_2)$ \\\addlinespace[1.0mm]
$\delta(r,b)$ & metric distance between $r$ and $b$ \\\addlinespace[1.0mm]
\multirow{2}{*}{$\delta (r,b,d,\vec{m})$} & aggregated metric distance from $r$ \\
 & $\text{ }$  via $b$ to $d$ with advertised $\vec{m}$ \\\addlinespace[1.0mm]
$cp(t)$ & capacity of link type $t\in T$ \\\addlinespace[1.0mm]
$cp(\ell)$ & capacity of link $\ell\in L$ \\\addlinespace[1.0mm]
$cost(x)$ & cost associated with entity $x$ \\\addlinespace[1.0mm]
\multirow{2}{*}{$dm(p)$} & traffic demand of the flow \\
 & $\text{ }$ that corresponds to path $p$ \\\addlinespace[1.0mm]
\midrule
\textbf{Boolean} & \multicolumn{1}{c}{\textbf{Meaning}} \\
\midrule
\multirow{2}{*}{$cr(f,p)$} & flow $f$ corresponds to path $p$ (i.e., both \\
 & $\text{ }$ have the same src. and dest. nodes) \\\addlinespace[1.0mm]
$dst(p,d)$ & $d\in N$ is the destination of path $p$ \\\addlinespace[1.0mm]
$tr(p,\ell)$ & path $p$ traverses link $\ell$ \\\addlinespace[1.0mm]
\multirow{2}{*}{$cons(p,\vec{m})$} & using path $p$ is consistent with \\
 & $\text{ }$ the advertisement of $\vec{m}$ \\\addlinespace[1.0mm]
\bottomrule
\end{tabular}\normalsize
\caption{Summary of Notation}\label{netwsymbls}
\end{center}\end{table}

SDN Partitioning allows to advertise topology information customized per sub-domain, as illustrated in Figure~\ref{view}: the original topology and its partitioning (shown in Subfig.~a) is not exposed to the nodes in sub-domain $k$ (i.e., nodes 1, 2, and 3). Instead, the customized view provided to these nodes pretends that both border nodes $a$ and $b$ have direct links to all other nodes (like shown in Subfig.~b). This way, the exit node for each inter-sub-domain flow can be determined on a per-destination basis simply by setting the OSPF link weights of the virtual connections. For instance, setting $x$ and $y$ determines how traffic for $d$ exits sub-domain $k$. Please note that the OSPF nodes 1, 2, and 3 are not aware of the fact that they form a sub-domain, and believe that the border SDN nodes $a$ and $b$ are regular OSPF neighbors.

\par
\begin{figure}[t] \center
\includegraphics[width=8.5cm]{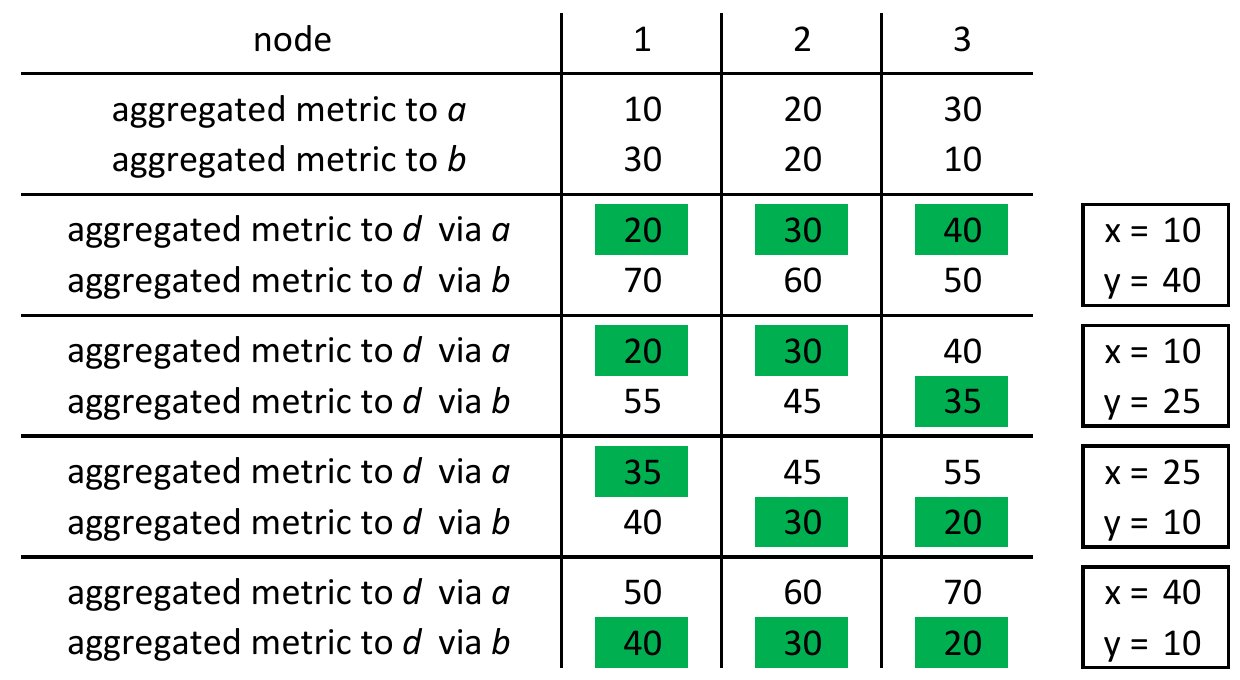}
\caption{Determination of exit nodes for flows to destination node $d$ from nodes 1, 2, and 3 in Figure~\ref{view}b, depending on the advertised link metrics $x$ and $y$.}
\label{LSA_sets} \end{figure}

 The expressiveness of routing in SDN-partitioned networks is constraint by the requirement that the usage of routing paths must be consistent with the advertised link metrics. More precisely, the routes of all flows, which start at the OSPF nodes $r_1\ldots r_\alpha$ of the same sub-domain and that are destined for the same sub-domain-external destination $d$, depend on the metrics $m(b_1,d) \ldots m(b_\beta,d)$ in the set of LSAs that has been advertised for $d$ by the border (SDN) routers $b_1\ldots b_\beta$ of that sub-domain. It can be seen in Figure~\ref{view} that traffic flows, which enter the network through a node of sub-domain $k$ and exit the network, for example, through node $d$, leave sub-domain $k$ either via node $a$ or $b$, depending on the aggregated OSPF link metrics to the border nodes plus the metric $m(a,d)$ advertised by $a$ and $m(b,d)$ advertised by $b$ (denoted as $x$ and $y$ in Figure~\ref{view}b) for the virtual links to $d$. If we assume that all links inside the sub-domain have an identical metric of 10, we can see all four possible routing scenarios in Figure~\ref{LSA_sets}. A field is marked green, if it represents the least cost path for a set of cost metrics. To give a counterexample for an impossible route combination, consider in Figure~\ref{view}b the two routes $1\rightarrow 2\rightarrow 3\rightarrow b\rightarrow d$ and $3\rightarrow 2\rightarrow 1\rightarrow a\rightarrow d$. There is no set of metrics $(m(a,d),m(b,d))$ that can be advertised by border nodes $a$ and $b$ that would allow this combination of routes, as the use of path $1\rightarrow 2\rightarrow 3\rightarrow b\rightarrow d$
presupposes
\begin{equation*}
\begin{split}
m(1,2)+ & m(2,3)+m(3,b)+m(b,d)<m(1,a)+m(a,d) \\
& \Rightarrow m(b,d) < m(a,d)
\end{split}
\end{equation*}
and the use of path $3\rightarrow 2\rightarrow 1\rightarrow a\rightarrow d$ presupposes
\begin{equation*}
\begin{split}
m(3,2)+ & m(2,1)+m(1,a)+m(a,d)<m(3,b)+m(b,d) \\
& \Rightarrow m(a,d) < m(b,d)
\end{split}
\end{equation*}
which is a contradiction. Consequently, the number of actually available route combinations is assumably less than the number of possible flow / exit node combinations.

\vspace{1mm}\noindent \textbf{Uniqueness:} A network with a set of nodes $N$ contains a set of  OSPF nodes $R$ and a set of SDN border nodes $B$ with $N = R \cup B$. A network partitioned into $K$ sub-domains is denoted as $K$-partitioned, and the sub-domains are ordered and numbered $1, 2, \ldots K$ in a unique fashion. Also, each node $n\in N$ has a network-wide unique identifier. Each node $r\in R$ belongs to a single sub-domain, and there is a subset $R_k$ of OSPF nodes for each sub-domain $k$. The $|R_k| = \alpha$ nodes in $R_k$ are ordered $R=(r_1 \ldots r_\alpha)$ in a unique fashion (e.g., lexicographically regarding their ID). Each (SDN) border node $b\in B$ belongs to multiple sub-domains, but for each sub-domain $k$, the order of its $\beta$ border nodes $B_k=(b_1 \ldots b_\beta)$ is unique as well. Furthermore, we denote $N_k=R_k\cup B_k$ as the set of all (i.e., domain-internal and border) nodes of sub-domain $k$ and $\overline{N}_k=N \backslash N_k$ as the set of all nodes external to sub-domain $k$.

\vspace{1mm}\noindent \textbf{Distance:} Each link $(n_1,n_2)$ is assigned an integer link metric value $m(n_1,n_2)$ and $m_k(n_1,n_2)$ denotes a dynamic metric (i.e., one that can be modified by the central controller) which is advertised in sub-domain $k$. A metric distance $\delta(r_1,b)$ is defined as the summation of link metrics $m(r_1,r_2)+m(r_2,r_3)\ldots +m(r_n,b)$ along the $n$ hop least cost path between an OSPF node $r_1$ and an SDN border node $b$ of the same sub-domain. All distances solely depend on the initial configuration of link metrics at OSPF nodes, and we assume that the configuration of those link metrics is static. We accordingly assume that distances are static for all sub-domains. Furthermore, we assume that least cost paths are unique (i.e., there exists always exactly one least cost path between any two nodes in the same sub-domain with metric distance $\delta$) and no mechanisms like Equal Cost Multi Path are in use. Finally, as a border node $b$ belongs to multiple sub-domains, dynamic metrics for a sub-domain-external destination $d$ are advertised on a per-sub-domain basis as $m(b,d,k)$ and additionally indicated by the sub-domain to which they belong.

\vspace{1mm}\noindent \textbf{Metric vector:} A set of link metrics that can be advertised by the $\beta$ border nodes of $B_k$ for a destination $d\in\overline{N}_k$ is denoted as metric vector $\vec{m} = (m(b_1,d,k), m(b_2,d,k),\ldots m(b_\beta,d,k))$. The components of a metric vector are ordered according to the ordering of the correspondent border nodes in that sub-domain and the $i^\text{th}$ component $m(b_i,d,k)$ is denoted as $\vec{m}_i$. A metric vector is \emph{valid}, if the advertisement of its link metrics lead to a nonambiguous single path routing scenario (i.e., without resulting in multiple least cost paths between any pair of nodes). Two metric vectors $\vec{m}\equiv \vec{m}'$ are equivalent if they result in the same routing. In other words, if we alter $\vec{m}$ to $\vec{m}'$ such that the changes of its elements are small enough and few enough, the routing will not change and we call the vectors equivalent. The equivalence class $[\vec{m}]$ of a metric vector is the set of all metric vectors $\vec{m}'$ with $\vec{m}'\equiv \vec{m}$. As equivalence in routing means redundancy, we ignore all other elements of $[\vec{m}]$ other than a single representative element $\vec{m}$. The set of all valid, non-equivalent (i.e. representative) metric vectors in sub-domain $k$ is denoted $M_k$. Please note that we interchangeably use the terms metric vector and the less formal but more common \emph{LSA set}.

\vspace{1mm}\noindent \textbf{Exit vector:} We denote the metric distance from node $r$ via border node $b$ to destination $d$ with link metrics $\vec{m}$ advertised for $d$ as $\delta (r,b,d,\vec{m})$. The exit border node $e(r,d,\vec{m})\in B_k$ for packets from a node $r\in R_k$ to a destination $d\in\overline{N}_k$ can now be determined by the used metric vector $\vec{m}$ that was advertised for $d$ in sub-domain $k$ as follows:
\[
\forall k\in [1,K], \, \forall d\in\overline{N}_k, \, \forall r\in R_k, \, \forall \vec{m}\in M_k:
\]\[
e(r,d,\vec{m}) = b' \,\,\,\, \text{if} \,\,\,\, \delta (r,b',d,\vec{m}) = \min\{ \delta (r,b,d,\vec{m}) | b\in B_k \}
\]
The exit vector $\vec{e}(k,d,\vec{m})=(e(r_1,d,\vec{m}), \ldots ,e(r_\alpha,d,\vec{m}))$ in sub-domain $k$ for packets from the $\alpha$ nodes $r\in R_k$ to the external destination $d\in\overline{N}_k$ is defined as the set of used exit border nodes (in the order according to $R_k$). We can shorten the notation to $e(r,\vec{m})$ and $\vec{e}(k,\vec{m})$ respectively, as the actual destination is irrelevant for generic considerations on exit nodes and exit vectors. We denote $E_k=\{ \vec{e}(k,\vec{m}) | \vec{m}\in M_k \}$ as the set of all exit vectors in sub-domain $k$. Please note that the mapping $M_k\rightarrow E_k$ is bijective by definition, as each representative metric vector determines one unique routing scenario.

\vspace{1mm}\noindent \textbf{Quantity vector:} The subset $R_k(b,\vec{m}) \subseteq R_k$ contains all domain-internal nodes of sub-domain $k$ that use border node $b$ as exit in case $\vec{m}\in M_k$ was advertised in sub-domain $k$. If $\vec{e}(k,\vec{m})$ is given, we can determine \emph{how many} OSPF nodes of $R_k$ use a specific $b\in B_k$ as exit. We denote this measure as \emph{quantity vector}:
\[
\vec{q}(k,\vec{m}) = (|R_k(b_1,\vec{m})|,\ldots , |R_k(b_\beta,\vec{m})|)
\]
We denote the set of all quantity vectors in sub-domain $k$ as $Q_k=\{ \vec{q}(k,\vec{m}) | \vec{m}\in M_k \}$. The mapping $E_k \rightarrow Q_k$ is obviously surjective, because each element $\vec{q}(k,\vec{m})\in Q_k$ is by definition generated by one element $\vec{e}(k,\vec{m})\in E_k$. More interestingly, the mapping is also injective and thus has a unique inverse mapping $E_k \leftarrow Q_k$, as we can always uniquely determine an exit vector if the quantity vector is given, which in turn has practical relevance regarding the complexity of the computation of all $M_k$ of a network.

\vspace{1mm}\noindent \textbf{Proof of injectivity:} The injectivity of $E_k \rightarrow Q_k$ can be proven by contradiction, if we assume that there exists a quantity vector $\vec{q}(k,\vec{m})$ that can be generated by two different exit vectors $\vec{e}(k,\vec{m}),\, \vec{g}(k,\vec{m'})$ with $\vec{m}\not\equiv \vec{m'}$, and thus $\vec{e} \neq \vec{g}$: A quantity vector is determined by an exit vector by \emph{counting} the occurrence of border nodes in it. It follows that if $\vec{e} \neq \vec{g}$ transform into the same quantity vector, $\vec{g}$ has to be a permutation of $\vec{e}$. In other words, the number of occurrences of each border node is the same in $\vec{e}$ and $\vec{g}$, and only their inner ordering is different. Such a permutation implies that a subset $X\in R_k$ of OSPF nodes has \emph{swapped} their exit nodes. Exit swapping, however, can in general be described as some OSPF node $r_1$ initially using border node $b_1$ and some other OSPF node $r_2$ using $b_2$ as exit with $m(b_1),\, m(b_2)$ advertised by $b_1,\, b_2$, or more formal:
\[
\delta (r_1,b_1) + m(b_1) < \delta (r_1,b_2) + m(b_2)
\]
and 
\[
\delta (r_2,b_2) + m(b_2) < \delta (r_2,b_1) + m(b_1)
\]
This can be reformulated to
\[
c_1 < m(b_1)-m(b_2) < c_2 
\]
with constant values
\[
c_1=\delta (r_2,b_2)-\delta (r_2,b_1) < c_2=\delta (r_1,b_2)-\delta (r_1,b_1)\tag{$\bigstar$}
\]
The different metrics $m'(b_1),\, m'(b_2)$ must now lead to the swapping of exit nodes, i.e., to
\[
c_1 > m'(b_1)-m'(b_2) > c_2
\]
which however implies $c_1>c_2$ and thus contradicts the relation marked with ($\bigstar$).\hfill\ensuremath{\square}

\subsection{LSA Generation Algorithm}\label{math_lsa}
The injectivity of $E_k \rightarrow Q_k$ has profound impact on the complexity of our algorithm that generates $M_k$ in our network, as the metric vectors can be identified based on quantity vectors rather than exit vectors. The search space of all metric vectors spans $\mathcal{O}(\beta^\alpha)$ (with $|R_k|=\alpha$ and $|B_k|=\beta$), while the search space of all quantity vectors only spans $\mathcal{O}(\binom{\beta+\alpha-1}{\alpha})$. A simple numerical example demonstrates the simplification: the upper bound on the number of metric vectors in a sub-domain with 4 SDN border nodes and 8 domain-internal OSPF nodes is 165 instead of 65536. The algorithm we developed to determine all metric vectors can be thought of as a tally counter, which always advances the element with the smallest index $i$ (i.e., the metric with the smallest index $i$ in $\vec{m}$) a single step (i.e., till the point where another OSPF node stops using the according border node $b_i$ as exit), until that element has reached the end (i.e., that metric is so large that $b_i$ is no longer the exit node for any OSPF node). In that event, that element is turned back to zero and the subsequent element is advanced, and so on. The algorithm works in detail as follows: The initial metric vector $\vec{m0}= (m_1,\ldots , m_\beta) \in M_k$ has values $m_1=0$ and $m_x=(x-1)\cdot (1+\Delta_k^\text{max})$ for all $2\leq x\leq\beta$ with $\Delta_k^\text{max}$ as the maximum difference between any two metric distances in sub-domain $k$:
\[
\Delta_k^\text{max} = \max\{\delta(r,b)-\delta(r',b') | r,r'\in R_k, \, b,b'\in B_k\}
\]
The quantity vector for this metric vector is $\vec{q}=(\alpha,0,\ldots,0)$, as all $|R_k| = \alpha$ OSPF nodes in $R_k$ will use the first border node as exit. (Please note our notation: the $n^\text{th}$ component of $\vec{a}$ is denoted $\vec{a}_n$ and the $i^\text{th}$ element of a set of vectors $A$ is denoted $\vec{ai}$.) The remaining metric vectors are generated iteratively, such that the $z^\text{th}$ metric vector $\vec{mz}$ is a copy of the $y^\text{th}$ metric vector $\vec{my}$ (where $z=y-1$) with its $i^\text{th}$ component (i.e., metric) increased by value $v$ as detailed in Algorithm~\ref{LSA_algo}.

\begin{algorithm}[h]\footnotesize
\KwIn{$R_k,B_k,\delta(r,b),\vec{my}$}
\KwOut{$\vec{mz}$}
\vspace{2mm}// Step 1: Determine the component index $i$\\
\For{$n:= \beta$ \KwTo $0$}{
	\If { $\vec{my}_{n}>0$}{
		$i := n$\;
	}
}
\If{$i = \beta$}{
	Terminate Algorithm!
}
\vspace{2mm}// Step 2: Determine value $v$\\
$v := \Delta_k^\text{max}$\;
\For{$n:=i+1$ \KwTo $\beta$}{
	\For{$q:=0$ \KwTo $\alpha$}{
		\If{$v > \delta(r_q,b_n)+\vec{my}_n$}{
			$v := \delta(r_q,b_n)+\vec{my}_n + 1$\;
		}
	}
}
\vspace{2mm}// Step 3: Generate $\vec{mz}$\\
$\vec{mz} := \vec{my}$\;
$\vec{mz}_i := \vec{mz}_i + v$\;
\For{$n:=i-1$ \KwTo $0$}{
	$\vec{mz}_n := 0$\;
}
\vspace{2mm}\caption{Generation of a Metric Vector}\label{LSA_algo}
\end{algorithm}

Among all representative metric vectors of $M_k$, this algorithm generates additionally a number of equivalent and a few non-valid ones, which have to be filtered out after the algorithm terminated.

\begin{figure}[h] \center
\includegraphics[width=7.7cm]{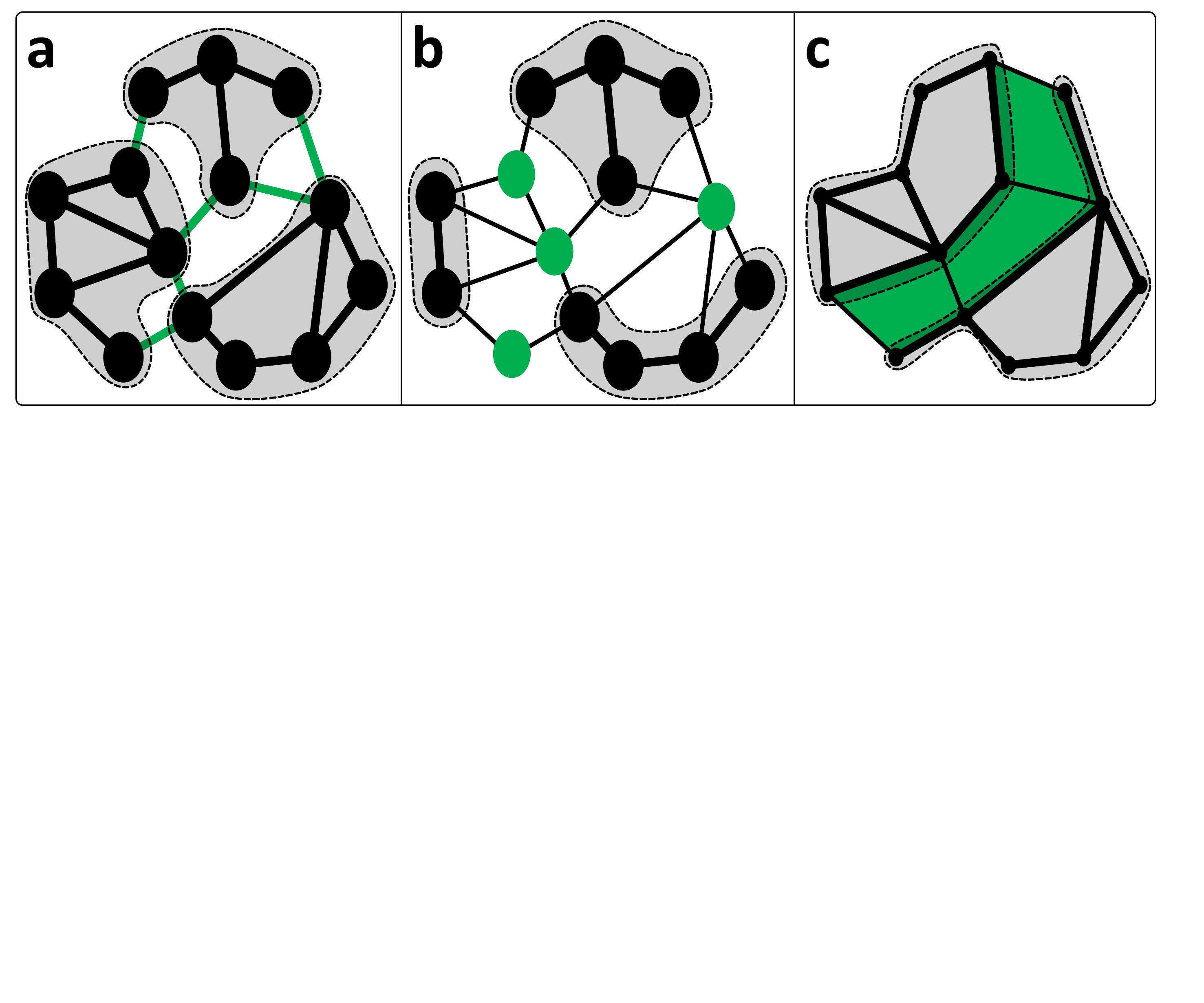}
\caption{A graph can be partitioned by removing a)~edges, b)~vertices, or c)~faces.}
\label{partitioning_methods} \end{figure}

\subsection{Network Partitioning}\label{math_vs}
\par Finding a good partitioning for a given topology is basically a graph theoretical discipline called graph partitioning, which has applications in many scientific and technical areas. Formally, the partitioning of a graph is a grouping of the graph's nodes into subgraphs, such that every node is included in one and only one of the subgraphs. More precisely, we define a $k$-way partitioning $P^k$ of graph $G=(N,L)$ as $P^k=\{T_1,T_2,\ldots T_k\}$ with $\bigcup_i T_i = N$ and $\bigcap_i T_i = \emptyset$. Graph partitioning is in general achieved by the (logical) deletion (or \emph{cut}) of some type of graph element, like shown in Figure~\ref{partitioning_methods}. In our case, as the application of graph partitioning is the segmentation of an OSPF domain into nonadjacent sub-domains by determining a set of SDN-enabled sub-domain border nodes, the graph theoretical problem we are dealing with is the search for a so-called \emph{vertex separator}. This is a set of nodes $\psi\subset N$, such that the removal of $\psi$ partitions $G$ into $k$ mutually unconnected subgraphs ${T_1,T_2,\ldots T_k}$. We assume that a graph is undirected, as we model IP networks, where all links are bidirectional.

\par For the most applications, determining a set of edges to be cut (the so called \emph{cut set}) is the natural approach for graph partitioning, which is also why the vast majority of algorithms that can be found in the literature tries to find a minimum cut set with some balancing constraint regarding the number (or aggregated weight) of nodes in the subgraphs~\cite{alpert}. A straightforward approach to find a good vertex separator is therefore to adapt one of these algorithms, simply by letting it derive a minimum cut set and to find a minimum adjacent node set (i.e., a minimum set of nodes, whose removal also removes the cut set), which however is not necessarily optimal. We therefore formulated an ILP model of network partitioning, which is based on the vertex separator ILP model published by Balas and de Souza in~\cite{Souza}. While the latter is limited to only two subgraphs, constrains an upper bound on the imbalance of the subgraph sizes, and has the objective of minimizing the vertex separator (i.e., the number of SDN border nodes in our application), we find our approach more applicable to our specific networking problem: It allows for the partitioning of a graph into an arbitrary number $K$ of subgraphs, constrains the number of SDN nodes, and has the objective to balance the sizes of all $K$ subgraphs. Our approach results in extremely balanced subgraph sizes. The fundamental idea remains however the same: Each node either belongs to a single connected subgraph or is a vertex separator (i.e., SDN) node, thus we can demand that a node belongs to the same subgraph as any of its direct neighbors, unless the node or the neighbor is an SDN node. This way, we force all node subsets to be pairwise unconnected and disjunct. We then balance the partitioning by putting a punishment cost on each subgraph that quadratically increases with its size, and then minimize the total cost over all subgraphs. We used the notation (if not already determined in Table~\ref{netwsymbls}) that is shown in Table~\ref{partitioning_vars}.

\begin{table}[t]\begin{center}\footnotesize
\begin{tabular}{ c l }
\toprule
\textbf{Real} & \multicolumn{1}{c}{\textbf{Meaning}} \\
\midrule
$\kappa(k)$ & utilization cost assigned to subgraph $k$ \\\addlinespace[1.0mm]
\midrule
\textbf{Integer} & \multicolumn{1}{c}{\textbf{Meaning}} \\
\midrule
$\varepsilon(k)$ & number of nodes in subgraph $k$ \\\addlinespace[1.0mm]
\midrule
\textbf{Boolean} & \multicolumn{1}{c}{\textbf{Meaning}} \\
\midrule
$\gamma (n, k)$ & node $n$ is part of subgraph $k$ \\\addlinespace[1.0mm]
$\mu (n)$ & node $n$ is a border node \\\addlinespace[1.0mm]
\bottomrule
\end{tabular}\normalsize
\caption{Variables for network partitioning}\label{partitioning_vars}
\end{center}\end{table}

\begin{figure}[t] \center
\includegraphics[width=\columnwidth]{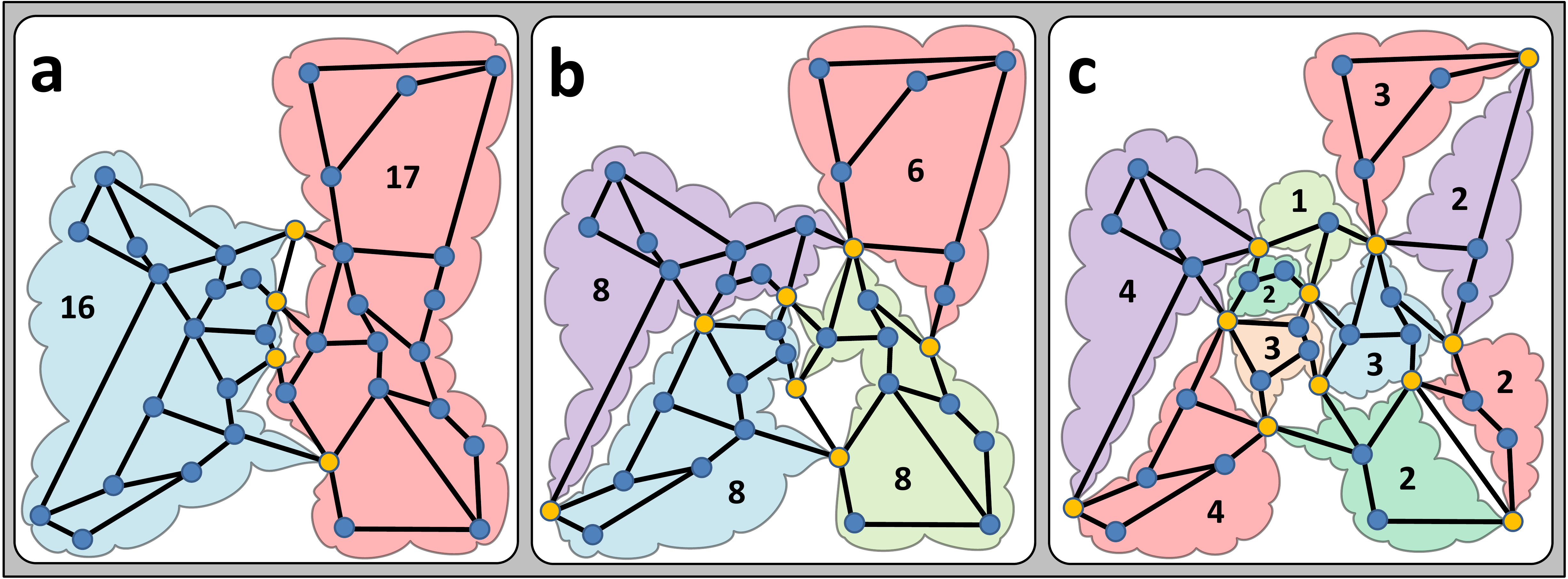}
\caption{Cost266 topology partitioned into 2,4, and 10 sub-domains.}
\label{cost266} \end{figure}

Our objective function is defined as
\[
\text{Minimize } \, \sum_{k\in[1,K]} \kappa(k)
\]

\begin{figure}[t] \center
\includegraphics[width=7.9cm]{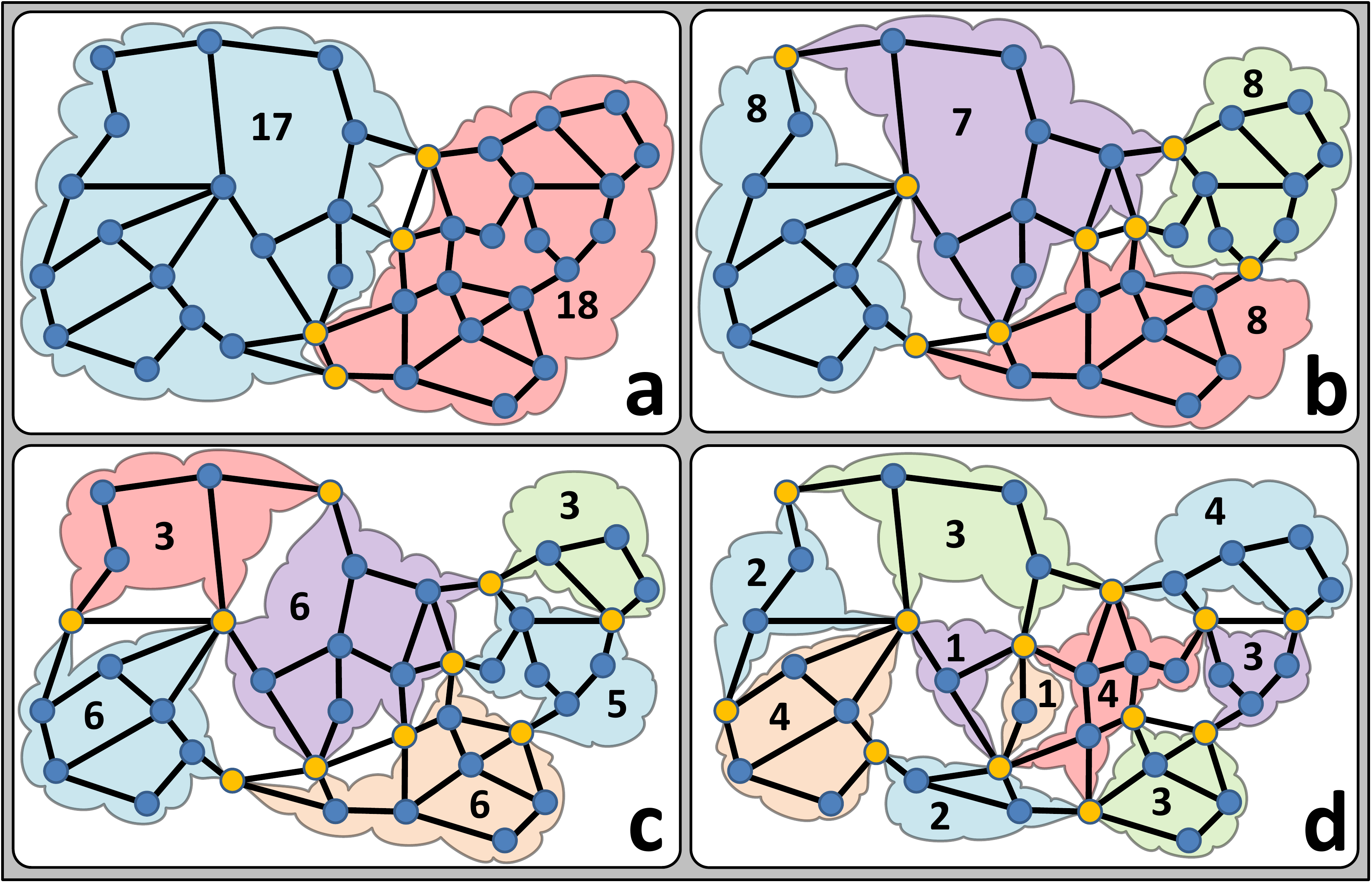}
\caption{Janos-US-CA topology partitioned into 2, 4, 6, and 10 sub-domains.}
\label{janos} \end{figure}

\begin{figure}[t] \center
\includegraphics[width=\columnwidth]{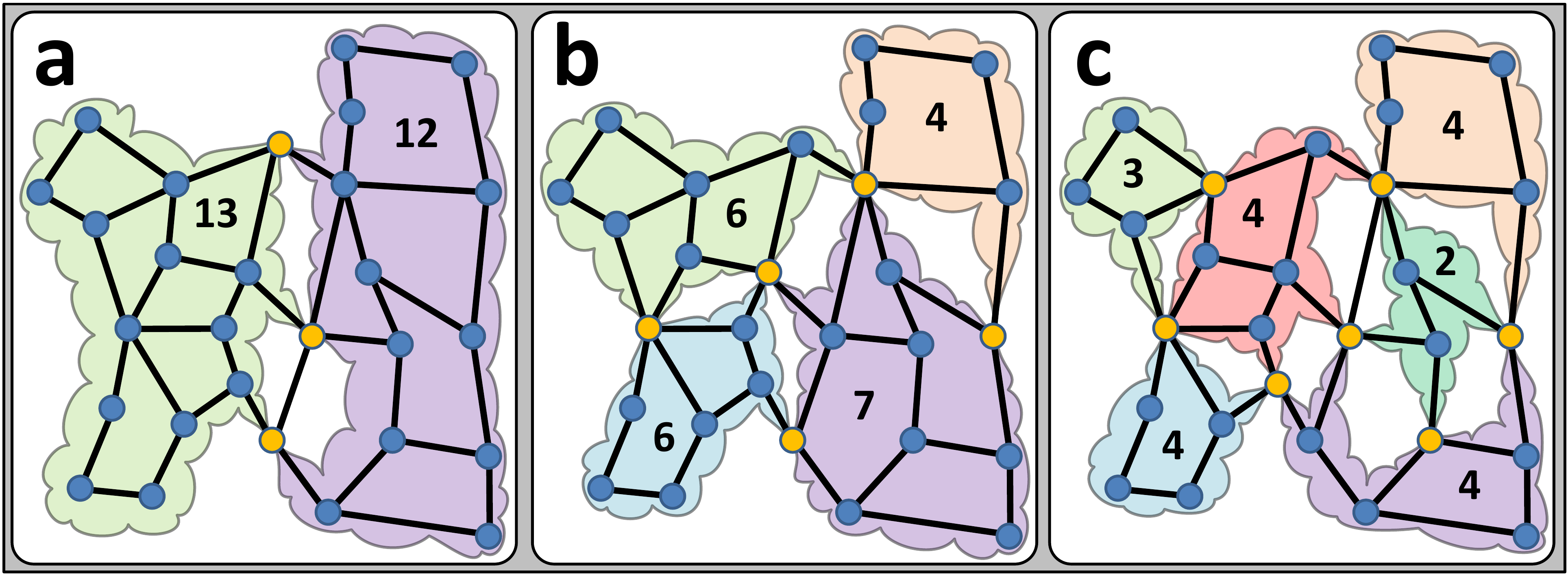}
\caption{Nobel-EU topology partitioned into 2, 4, and 6 sub-domains.}
\label{nobel} \end{figure}

This objective requires a constraint that determines the cost per sub-domain, such that it increases, as described above, quadratically with its size. However, as a quadratically increasing function can not be used in linear optimization, we have to determine the cost increase with a piecewise linear inequation over a set $Y$ of linear functions:
\[
\forall y\in Y, \, \forall k\in [1,K]: \,\, \kappa(k) \geq y_a\cdot \varepsilon(k) +y_b
\]
We here apply the linear cost functions $y\in Y$ similar to the first constraint of the traffic engineering model in Subsection~\ref{math_te} as lower bound on the cost. Now we need to determine the size of sub-domain $k$ simply by counting all its nodes:
\[
\forall k\in [1,K]: \,\, \varepsilon(k) = \sum_{n\in N} \gamma(n,k)
\]
This in turn requires a constraint that assures that each node is either an SDN node or it belongs to a single sub-domain:
\[
\forall n\in N: \,\,
\mu(n) + \sum_{k\in[1,K]} \gamma(n,k) = 1
\]
The next constraint assures that neighboring nodes belong to the same sub-domain:
\[
\forall k\in [1,K], \, \forall i,j\in N \text{ with } (i,j) \in L:
\]
\[
\gamma(i,k) - \mu(i) \geq \gamma(j,k)
\]
This constraint assures that node $j$ belongs to sub-domain $k$ only if its neighbor $i$ belongs to the same sub-domain or is an SDN node. Finally, the next constraint limits the total number of nodes that can be SDN-enabled to the upper bound $\text{MAX}$:
\[
\sum_{n\in N} \mu(n) \leq \text{MAX}
\]
The model solves quickly for networks with $|N|<50$, which allows to fine tune parameters $K$, $\text{MAX}$, and to additionally introduce lower and upper bounds $lb \leq \varepsilon \leq ub$ on the subgraph size. The partitioning of various samples topologies is illustrated in Figures~\ref{cost266}, ~\ref{janos}, and ~\ref{nobel}.

\section{Network Optimizations}\label{optimization}
\subsection{Capacity Dimensioning}\label{math_ne}
Capacity dimensioning is an important network management task, where the links of a network are re-dimensioned by the operator to accommodate the network capacity to the changing traffic demands for the next planning period. Shortest path routing would always result in minimum capacity requirements, if link capacities could be commissioned in arbitrary sizes, but link capacities are available in our assumed transport network architecture only in fixed granularities (i.e., 10~Gbps, 40~Gbps, and 100~Gbps) and with each capacity type there is a cost associated for the required devices in the optical layer (i.e., transponders, line cards, energy, etc.). Capacity planning is trivial in OSPF networks, when the routing remains unchanged and the traffic demand is known for the next planning period: if the estimated traffic load on a link exceeds a certain utilization threshold, the link is provisioned with the next larger capacity granularity. However, more sophisticated control planes allow to steer the routing to avoid underutilized links (and thus capacity wasting), which for instance occur in case of an estimated demand of 11~Gbps in the next planning period on a link that was previously provisioned with 10~Gbps capacity: This link requires an upgrade to the next capacity level (i.e., 40~Gbps) and will exhibit a poor utilization of less than 28\% during the next planning period. A sophisticated control plane thus routes the traffic flows such that poorly utilized links are avoided, which allows to reduce the total capacity requirements to the actual demand. We will explain in this subsection the mathematical model for capacity planning in SDN-partitioned OSPF networks. The objective of this model is to determine the traffic routing that requires the minimum total cost for the provisioned link capacities, while it suffices all architectural constraints. In addition to the parameters summarized in Table~\ref{netwsymbls} we use the variables in Table~\ref{NE_vars}.

\begin{table}[t]\begin{center}\footnotesize
\begin{tabular}{ c l }
\toprule
\textbf{Boolean} & \multicolumn{1}{c}{\textbf{Meaning}} \\
\midrule
$\psi(\ell,t)$ & capacity type $t$ is used on link $\ell$ \\\addlinespace[1.0mm]
\multirow{2}{*}{$\varphi (\vec{m}, d)$} & metric vector $\vec{m}$ is advertised \\
 & $\text{ }$ for destination $d$ \\\addlinespace[1.0mm]
$\rho(p)$ & path $p$ is used \\\addlinespace[1.0mm]
\bottomrule
\end{tabular}\normalsize
\caption{Variables for capacity dimensioning}\label{NE_vars}
\end{center}\end{table}

The objective function of this model minimizes the summation of the cost for the commissioned capacity types of all links in the network:
\begin{equation*}
\text{Minimize}\,\,\,\,\,\, \sum_{\ell\in L} \sum_{t\in T} \psi(\ell , t) \cdot cost(t)
\end{equation*}
subject to all constraints explained below. The first constraint limits the amount of traffic on a link to the commissioned link capacity in consideration of the maximum allowed link utilization $u_\text{max}$:
\begin{equation*}
\forall \ell\in L: \,\,\sum_{p\in P} \rho(p) \cdot tr(p, l) \cdot dm(p) 
\leq \sum_{t\in T} \psi(\ell , t) \cdot cp(t) \cdot u_\text{max}
\end{equation*}
We need three additional constraints on routing and the use of metric vectors, which are also required in the subsequent mathematical models (and we therefore assigned reference numbers). The first one assures that exactly one path will be used for each flow $f$ (which corresponds to a specific source-destination pair of nodes) in the network, i.e., 
\begin{equation}\label{eq1}
\forall f\in F: \,\, \sum_{p\in P} \rho(p) \cdot cr(f,p)=1
\end{equation}
Please note that all possible routing paths are precomputed according to the scheme's inherent working principle and subsumed in $P$, in order to reduce the complexity of this ILP model. The second routing constraint guarantees that the choice of paths to be used is segment-wise (i.e., per sub-domain) consistent with the link-weight-based routing behavior of OSPF routers:
\begin{multline}\label{eq2}
\forall k\in [1,K], \,\, \forall p\in P, \,\,  \forall d\in\overline{N}_k \,\, \text{with} \,\, dst(p,d):\\
\rho(p) \leq \sum_{\vec{m}\in M_k} \varphi (\vec{m}, d) \cdot cons(p,\vec{m}) \,\,\,\,\,\,\,\,\,
\end{multline}
The precomputed parameter $cons(p,\vec{m})$ is 1 only if the advertisement of link metrics in $\vec{m}$ in sub-domain $k$ for destination $d$ lead to an OSPF forwarding behavior such that the exit border node in $k$ is contained in path $p$, and 0 otherwise.

Finally, we need a constraint to assure that in each sub-domain exactly one metric vector is advertised for each sub-domain-external destination:
\begin{equation}\label{eq3}
\forall k\in [1,K], \,\, \forall d\in \overline{N}_k:   \,\,\, \sum_{\vec{m} \in M_k} \varphi (\vec{m},d) =1
\end{equation}

\subsection{Traffic Engineering}\label{math_te}
Traffic engineering is an important network management task, where the routing of traffic flows is changed to balance the traffic load over all network links (i.e., to even up their utilization). We will now explain the mathematical model for traffic engineering in SDN-partitioned OSPF networks. In addition to the parameters summarized in Table~\ref{netwsymbls} we use the variables in Table~\ref{TE_vars}.

\begin{table}[t]\begin{center}\footnotesize
\begin{tabular}{ c l }
\toprule
\textbf{Real} & \multicolumn{1}{c}{\textbf{Meaning}} \\
\midrule
$\kappa(\ell)$ & utilization cost assigned to link $\ell$ \\\addlinespace[1.0mm]
\midrule
\textbf{Boolean} & \multicolumn{1}{c}{\textbf{Meaning}} \\
\midrule
\multirow{2}{*}{$\varphi (\vec{m}, d)$} & metric vector $\vec{m}$ is advertised \\
 & $\text{ }$ for destination $d$ \\\addlinespace[1.0mm]
$\rho(p)$ & path $p$ is used \\\addlinespace[1.0mm]
\bottomrule
\end{tabular}\normalsize
\caption{Variables for load balancing}\label{TE_vars}
\end{center}\end{table}

The objective function minimizes the summation of the utilization cost of all links in the network:
\begin{equation*}
\text{Minimize}\,\,\,\,\,\, \sum_{\ell\in L} \kappa(\ell)
\end{equation*}
subject to all constraints explained below. The first constraint assigns the utilization cost to all links based on the utilization of the link and the linear cost functions:
\[\ell\in L, \,\, \forall y\in Y:\]
\[
\kappa(\ell) \geq y_a \cdot \Bigg( \sum_{f\in F} \bigg( \frac{dm(f)}{cp(\ell)} \sum_{p\in P} \rho(p)\cdot tr(p,\ell) \cdot cr(f,p) \bigg) \Bigg) + y_b
\]
On the right-hand side of this inequation, the summation over $p\in P$ is just a boolean indicating whether there is a path used for the considered flow $f$ that traverses the considered link $\ell$. If this is the case, we add the demand of the flow divided by the capacity of the link to the utilization term, which is the summation over all flows $f\in F$ (i.e., the term in the largest brackets). Finally, the assignment of utilization cost $\kappa$ to the link depends on all linear cost functions $y\in Y$, i.e., a particular $y$ generates a lower bound on the cost by multiplying the utilization with a constant $y_a$ and adding another constant $y_b$. Additionally, we need only three more constraints to complete the model, which are however the same as Eq.\ref{eq1}, Eq.\ref{eq2}, Eq.\ref{eq3} on routing and metric vectors like in the previous model.

\subsection{Failure Recovery}\label{math_fr}
Failure recovery is the ability of a control plane to react on failures in the network with an alternative valid routing (i.e., without routing loops and black holes) to avoid further traffic loss. We will now explain the mathematical model for failure recovery in SDN-partitioned OSPF networks. Its objective is to react on link failures with the fewest possible routing updates, while at the same time avoiding over-utilization on any link. The two objectives are conflicting, which requires to trade routing stability off against balanced link utilizations. This model resembles the previous one for the most part, because the mechanism basically load balances again, but here with a punishment on new link metric advertisements to keep the routing in the OSPF part of the hybrid control plane as stable as possible. However, routing changes through updates in the flow tables of the SDN border nodes are considered as invisible for OSPF and therefore do not provoke routing recomputation in legacy nodes. Therefore, our objective function resembles the one of the previous model, and additionally considers only the (stability) cost of metric changes (and not routing changes in general). We assume that in case of a failure on link $\ell$ in sub-domain $k$ failure recovery is carried out as follows: all OSPF paths in sub-domain $k$ that contain link $\ell$ are recomputed by the sub-domain's OSPF nodes (which is conform to regular OSPF operation), which in turn changes the affected $\delta(r,b)$, and thus the mapping of the pre-computed metric vectors $\vec{m}\in M_k$ to the exit vectors $\vec{e}\in E_k$. All these parameters and the entire set of available routing paths $P$ have to be recomputed as well for the here presented recovery model. The objective function is a minimization of the link utilization cost (like in the previous model) plus the (stability) cost for link metrics that are changed through the recovery process:
\begin{equation*}
\text{Minimize:}\,\, \sum_{\ell\in L} \kappa(\ell) + \sum_{\varphi} \varphi \cdot cost(\varphi)
\end{equation*}
We define the here used cost parameter $cost(\varphi)$ for the advertisement of metric vectors as the number of individual metric changes multiplied by a predefined punishment cost. In other words, all components of a metric vector $\vec{m}$ advertised for a specific destination through failure recovery, which are different from the metric vector advertised for the same destination before the link failure are counted and multiplied with some punishment cost. This punishment cost is then summarized over all metric advertisement variables $\varphi$ and added to the objective function.

\section{Numerical Analysis}\label{results}
\begin{figure}[t] \center
\includegraphics[width=\columnwidth]{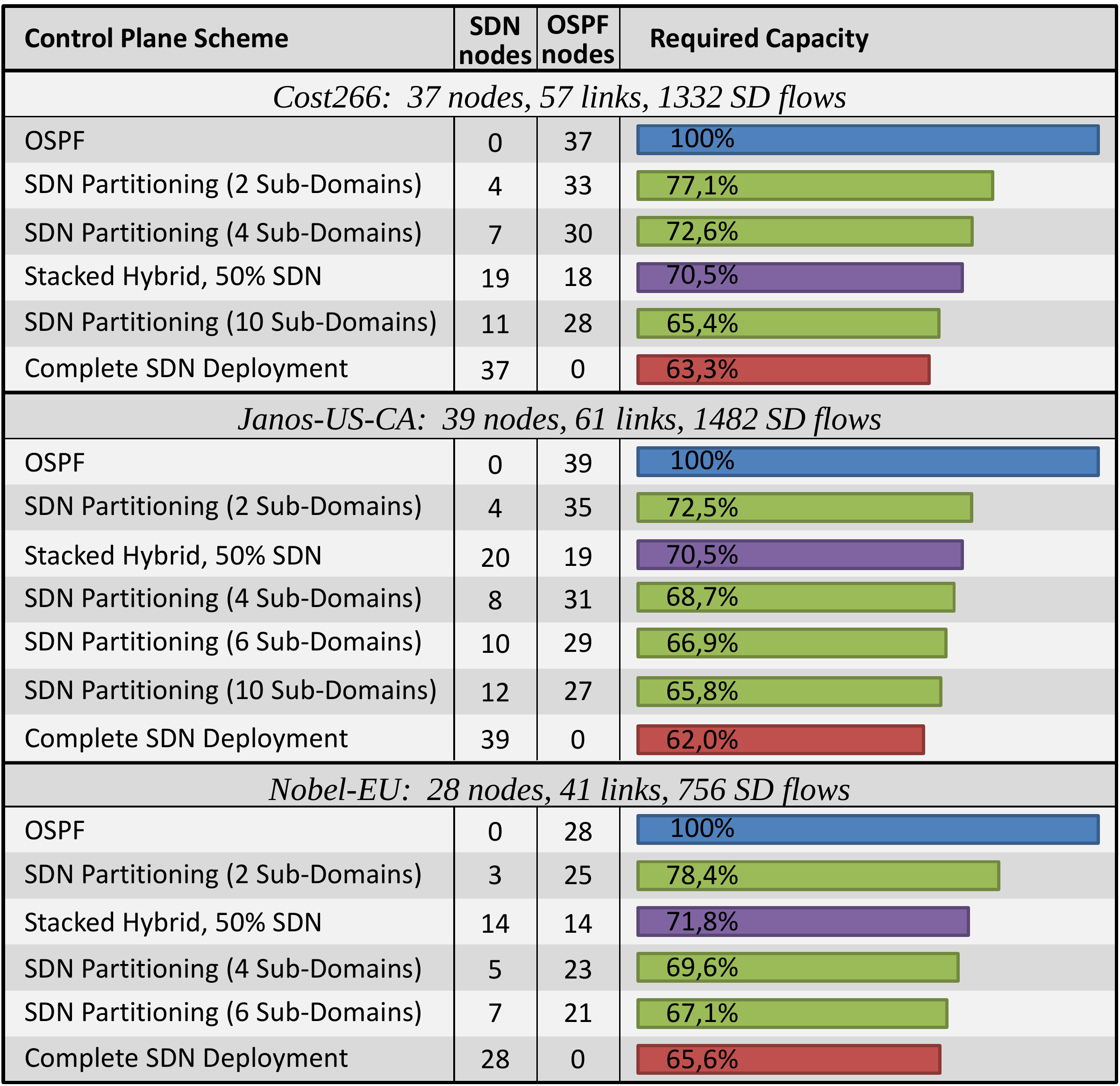}
\caption{Required link capacities in the different topologies depending on the used control plane scheme (normalized to OSPF requirements).}
\label{results_NE} \end{figure}

For our performance analysis, we used the Cost266, the Janos-US-CA, and the Nobel-EU topologies from the SNDlib library~\cite{sndlib}. Figure~\ref{results_NE} shows the results of capacity planning in relation to the requirements of an OSPF-controlled network. OSPF is taken as worst case, as no load balancing is applied, whereas all other operational schemes allow to optimize the routing to improve resource utilization, which in turn allows to reduce the link capacities. Figure~\ref{results_NE} compares the capacity requirements in the three different network topologies depending on the used control scheme, whereas the SDN Partitioning results are furthermore classified depending on the actually applied partitioning into sub-domains of the initial topology. The Cost266 topology and the used partitionings into 2, 4, and 10 sub-domains is depicted in Figure~\ref{cost266}, the Janos-US-CA topology with partitionings into 2, 4, 6, and 10 sub-domains is depicted in Figure~\ref{janos}, and the Nobel-EU topology with partitionings into 2, 4, and 6 sub-domains is depicted in Figure~\ref{nobel}. We compare the performance of SDN Partitioning (using the optimization model introduced in Subsect.~\ref{math_ne}) furthermore with full SDN deployment and the most commonly used hybrid SDN/OSPF control plane scheme (denoted as the ``stacked'' hybrid scheme), where all nodes participate in OSPF and hybrid nodes can additionally be configured dynamically with high priority routing rules. For this scheme we assumed that (at least) 50\% of all nodes are SDN-enabled and the optimal location of these nodes was determined based on the location optimization method in~\cite{hybrid_2}. The actual number of SDN-enabled and legacy OSPF nodes is given in the second and third column of Figure~\ref{results_NE}.

The evaluation of capacity requirements were carried out as follows: we used the initially unpartitioned network for the ``OSPF'' case, assigned uniform link metrics, and determined the OSPF least cost paths, which resulted in minimum hop count routing. We then assigned uniformly distributed traffic demands to all source-destination pairs in the network and rescaled them all with the same scaling factor, such that the maximum link load is set to 80~Gbit/s. We assigned  minimum capacities to all links such that no link loads exceeds 80\% of the link's capacity. We assumed that link capacities are available in granularities 10~Gbit/s, 40~Gbit/s, and 100~Gbit/s. All other results in Figure~\ref{results_NE} show the minimum capacity requirements of the according control plane scheme after the routing has been optimized under the schemes' individual routing constraints. The first noticeable characteristic of this result is that \emph{all} evaluated schemes are able to save considerable amounts of link capacity compared to OSPF. This was however to be expected, as link utilization is not considered in the routing algorithm of OSPF, which thus can lead to significant capacity wasting. Packets are solely routed via shortest (i.e., least metric cost) paths, which may result in link loads that slightly exceed link capacity granularities (e.g., 11~Gbit/s traffic load necessitates 40~Gbit/s link capacity). More remarkable is however, to what extent SDN Partitioning can keep up with or even outperform the 50\% and 100\% SDN deployment. It can be seen in the figure that the partitioning with the most sub-domains is relatively close to the result of full SDN deployment in each tested topology. We generally conclude that a migration to SDN-enabled devices beyond the requirements of SDN Partitioning (with small sub-domains) can not result in significant further capacity savings. Another remarkable outcome of this evaluation is the fact that SDN Partitioning outperforms stacked hybrid SDN/OSPF operation with only a fraction of the required SDN-enabled nodes. It can finally be seen that even the partitioning into only two sub-domains can considerably improve resource utilization compared to plain OSPF, while the number of required SDN nodes is notably low.

\begin{figure}[t] \center
\includegraphics[width=\columnwidth]{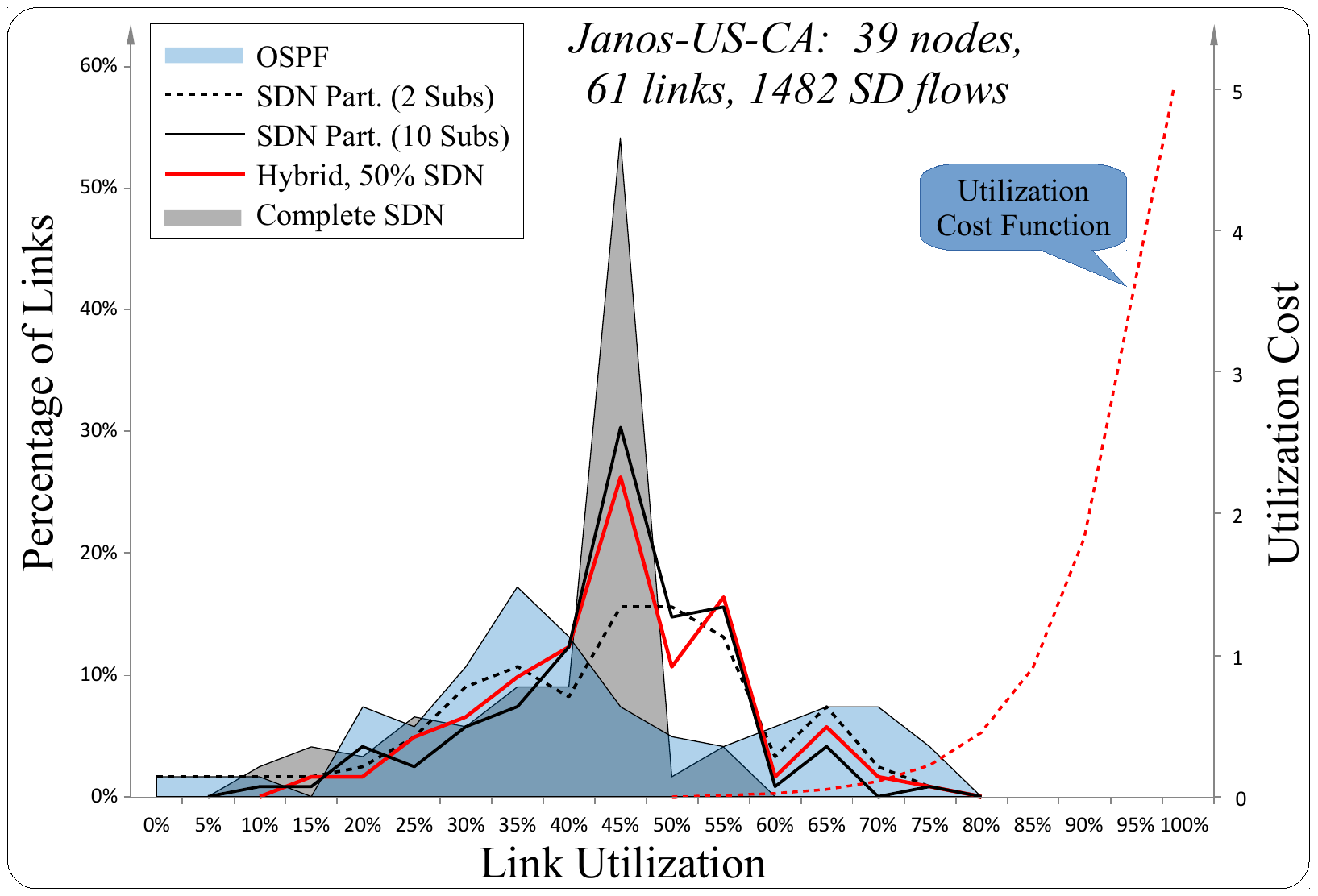}
\caption{Histograms of link utilization of the different control plane schemes.}
\label{results_TE} \end{figure}

Figure~\ref{results_TE} shows the performance of load balancing in the Janos-US-CA topology in the form of histograms of link utilization, defined as the frequency of the occurrence of a particular link utilization value. The according experiments were carried out as follows: As initial scenario we used the traffic and link capacity values determined in the OSPF case of the previous experiment. The blue area in the figure depicts how OSPF utilizes the deployed links, which covers a wide range. This was again used as worst case result without any load balancing. We then used the routing optimization model detailed in Subsection~\ref{math_te} to balance the link loads such that the occurrence of higher utilization degrees is less frequent. We again used the identical objective for the 50\% (stacked hybrid) and the complete SDN deployment. The utilization cost function is superimposed in the figure (shown as the dotted red ``Cost'' plot). The optimality bound for load balancing is the case where all links are exactly equally utilized, which would result in a histogram with a single peak with 100\% of links. Indeed, the histogram of full SDN deployment (plotted as gray area) exhibits a strong peak (54.1\% of all links) right below 50\% link utilization, which is exactly the upper bound of the ``zero cost zone'' (i.e., links with $\leq 50\%$ utilization cause zero cost in the routing optimization model). This result can be considered as the best case result for load balancing when routing is not constrained. Figure~\ref{results_TE} also shows the histograms for SDN Partitioning with 2 (dotted black line) and 10 (solid black line) sub-domains. (Please note that due to clarity of this illustration we omitted the plots for SDN Partitioning with 4 and 6 sub-domains, which however -- if plotted in the same figure -- would smoothly integrate between the plots for 2 and 10 sub-domains.) SDN Partitioning with 10 sub-domains allows for extensive routing control resulting in load balancing performance close to full SDN deployment, which indicates that a relatively small number of SDN-enabled routers (even compared to the stacked hybrid scheme with 50\% SDN-enabled nodes, plotted as solid red line) can enable almost full traffic engineering capabilities in a network. The figure also shows that SDN Partitioning with only 2 sub-domains can already considerably improve the load distribution compared to regular OSPF operation.

\begin{figure}[t] \center
\includegraphics[width=\columnwidth]{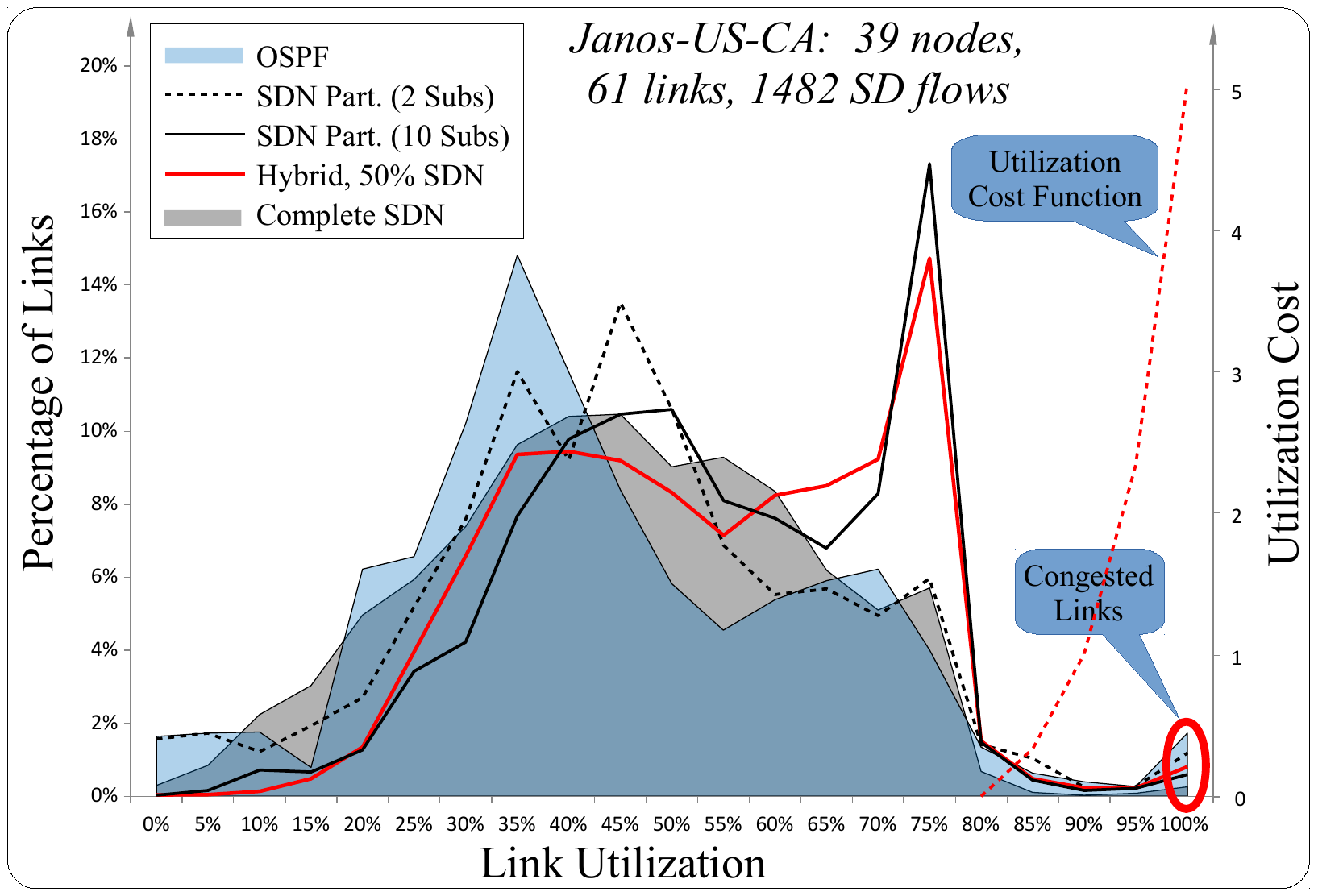}
\caption{Histograms of the averaged link utilization of the different control plane schemes after the occurrence of a link failure.}
\label{results_FR} \end{figure}

Our final result is depicted in Figure~\ref{results_FR} and shows (like in the previous result in the form of link utilization histograms) to what extent the compared operational schemes can handle the occurrence of a sudden fiber cut in the network. This experiment was carried out based on the previous load-balancing scenario, where we simply deleted one link and reoptimized the routing. The results are averaged over all possible link failures in the network. For this experiment, we shifted the cost function to 80\% in order to avoid only over-utilized links (which can easily occur in case of a fiber cut), while trying to minimize the number of events\footnote{As an event in the OSPF part of the control plan we consider each routing recomputation in an OSPF router due to a new Link State Advertisement.} in the OSPF part of the control plane. In other words, routing optimization using the formulation in~\ref{math_fr} aims on fewest LSAs and re-routes flows only in case of over-utilized (or failed) links. The distribution of link utilizations in case no routing re-optimization is possible -- which is the case in OSPF -- is shown as the blue area. Here, OSPF only assures that all nodes compute valid new shortest path routes, which is based only on link metrics and completely ignores the traffic load. Consequently, OSPF leads to the largest number of congested links. (See Table~\ref{droppedpackets} for a numerical congestion comparison.) SDN Partitioning with 2 sub-domains (i.e., the control plane scheme with the weakest control on routing) is plotted as dotted black line and can already provide significant improvements compared to OSPF in terms of congestion. More sophisticated routing control is provided by SDN Partitioning with 10 sub-domains (solid black line) and stacked hybrid control with 50\% SDN-enabled routers (solid red line), which both lead to a significant peak at the lower cost bound at 80\% link utilization on the one hand, and decreased congestion on the other hand. Full SDN deployment (plotted as gray area) can almost completely avoid congestion in our experimental set up. Please note that it is common in operative IP networks either to provide protection (i.e., idle backup links) or to overprovision link capacities such that congestion is avoided in case of link failures, which is expensive in terms of required capacity.

In the same experiment we measured the amount of excess traffic (i.e., packets that are dropped due to overloaded links) \emph{after the control plane has finished all its routing reconfigurations} after a link failure. (Note that traffic loss \emph{during} routing reconfiguration is ignored in this table, as we assume that its duration is very short compared to the duration of the link failure.) These measurements are provided in Table~\ref{droppedpackets}, which additionally provides the average fraction of links that exhibit congestion and a measure on OSPF routing stability quantified by the number of actually performed routing recomputations in OSPF routers. Given our assumption of an initial link utilization threshold of 80\%, a link failure in such an ``economically'' dimensioned OSPF network leads to drastic service degradation, which -- on average -- already entail more than 0.7\% of all packets dropped due to congestion (which also implies increased packet delays). Even though packet loss can not be avoided completely under these assumptions, the table shows that improved routing control can reduce the amount of excess traffic load significantly.

\begin{table}[t]\begin{center}\footnotesize
\begin{tabular}{ l c c c}
\toprule
\multirow{2}{*}{\textbf{Operational Scheme}} & \textbf{Traffic} & \textbf{Cong.} & \textbf{OSPF} \\
                                             & \textbf{Loss}    & \textbf{Links} & \textbf{Reconf.} \\
\midrule
OSPF & $7.24\permil$ & 1.73\% & 78 \\\addlinespace[1.0mm]
SDN Partitioning (2 Subs) & $3.57\permil$ & 1.19\% & 430.4 \\\addlinespace[1.0mm]
Stacked Hybrid, 50\% SDN & $2.47\permil$ & 0.81\% & 78 \\\addlinespace[1.0mm]
SDN Partitioning (4 Subs) & $2.28\permil$ & 0.68\% & 110.5 \\\addlinespace[1.0mm]
SDN Partitioning (6 Subs) & $2.03\permil$ & 0.60\% & 45.8 \\\addlinespace[1.0mm]
SDN Partitioning (10 Subs) & $1.89\permil$ & 0.60\% & 24.7 \\\addlinespace[1.0mm]
Complete SDN Deployment & $0.82\permil$ & 0.26\% & 0 \\\addlinespace[1.0mm]
\bottomrule
\end{tabular}\normalsize
\caption{OSPF routing stability and average traffic loss through overutilized links after a sudden fiber cut in the Janos-US-CA topology.}\label{droppedpackets}
\end{center}\end{table}

Routing stability in the OSPF part of a hybrid control plane is an important aspect, as each new LSA received by an OSPF router triggers a recomputation of the routing and forwarding table. The last column of Table~\ref{droppedpackets} shows the average number of such OSPF reconfigurations for each examined control plane. Regular OSPF triggers exactly 78 of such events after any link failure, as the used topology has 39 nodes and both adjacent routers advertise the topology change through flooding to the entire routing domain. The stacked hybrid control plane behaves identical, as all hybrid nodes perform regular OSPF below their SDN layer. However, as soon as all legacy nodes are substituted with SDN nodes, the legacy protocol can finally be turned off completely, which consequently results in zero OSPF reconfigurations for the case of complete SDN deployment. It can be seen from the last column of Table~\ref{droppedpackets} that in case of SDN Partitioning the number of OSPF reconfigurations strongly depends on the number of sub-domains. Using this scheme with only two sub-domains provokes excessive use of OSPF reconfigurations, which however allows at least to half the amount of lost traffic compared to native OSPF operation. Due to the low number of SDN routers (4 out of 39) in this scenario, the capability to reroute traffic around congested areas (or failed links) by means of flow table updates from the central SDN controller is comparably limited. The only other method to change routing in SDN Partitioning is to change the SDN border nodes used as sub-domain exit on a per-destination base, which in this case is heavily used by the failure recovery process to reduce packet loss and link congestion. It can however also be seen from the table that OSPF's routing stability increases rapidly in SDN Partitioning with an increasing number of sub-domains (and thus SDN nodes).

\section{Conclusions}\label{conclusions}
The advantages of a hybrid SDN/OSPF control plane are broadly recognized in the networking community, as it promises the best of two worlds: programmability and agility of SDN and reliability and fault tolerance of OSPF. Moreover, such an architecture allows for a smooth and cost optimized migration to SDN. The common approach for such a control plane follows a ``ships-passing-in-the-night'' strategy, where the SDN part and the OSPF part are oblivious of what the other configures. We have argued that this is however not without issues, especially in terms of forwarding table sizes, routing convergence time, location of the SDN-enabled nodes, and in case of network failures. We have proposed SDN Partitioning in this paper as a new hybrid SDN/OSPF control plane that encapsulates OSPF into sub-domains and allows to steer the routing between sub-domains. We provided a new ILP-based algorithm for balanced vertex separators that can be used to equally partition a routing domain and demonstrated the correlation of the number of deployed SDN border nodes and sub-domain size. We have explained in detail all technical requirements and provided new mathematical models that take into account the specific routing constraints for common network management tasks, namely capacity planning, load balancing, and failure recovery. Finally, we numerically evaluated the performance of SDN Partitioning in comparison to OSPF operation, full SDN deployment, and hybrid SDN/OSPF (assuming a 50\% SDN deployment) without partitioning. Our results show that -- depending on the degree of partitioning -- SDN Partitioning provides network control capabilities between 50\% and full SDN deployment, but with relatively few SDN-enabled routers. Adjusting the sub-domain sizes allows to trade off the degree of dynamic control (and thus the performance of the evaluated management operations) against carefreeness and routing stability. This claim is confirmed by our numerical results: larger sub-domains provide less SDN control (due to more autonomous OSPF self-configuration), while smaller sub-domains increase the domination of the SDN control plane (and thus the performance of management operations that depend on dynamic routing control), but require a larger number of SDN-enabled nodes in the network. This also proves that SDN Partitioning provides a pragmatic and efficient migration path for network operators, as an initial partitioning into two sub-domains requires only a few SDN nodes. Sub-domains could iteratively be partitioned into smaller sub-domains in further migration steps, which would gradually increases the central control on routing for manageable investments in new equipment.

\section*{Acknowledgments}
This work has been supported by the German Federal Ministry of Education and Research (BMBF) under code 01BP12300A; EUREKA-Project SASER.


\end{document}